\documentstyle[12pt,epsfig,multirow]{article}

\newcommand{\be}{\begin{eqnarray}}
\newcommand{\ee}{\end{eqnarray}}

\def\nue{{\nu_e}}
\def\anue{{\bar\nu_e}}
\def\numu{{\nu_{\mu}}}
\def\anumu{{\bar\nu_{\mu}}}

\newcommand{\ms}{\Delta m^2_{21}}
\newcommand{\ma}{\Delta m^2_{31}}

\newcommand{\sss}{\sin^2 \theta_{12}}
\newcommand{\sch}{\sin^2 \theta_{13}}
\newcommand{\stch}{\sin^2 2\theta_{13}}

\newcommand{\sta}{\sin^22 \theta_{23}}
\newcommand{\mst}{\Delta m^2_{21}{\mbox {(true)}}}
\newcommand{\mat}{\Delta m^2_{31}{\mbox {(true)}}}
\newcommand{\sgnma}{$sgn(\Delta m^2_{31})$}

\newcommand{\ssst}{\sin^2 \theta_{12}{\mbox {(true)}}}

\newcommand{\stcht}{\sin^2 2\theta_{13}{\mbox {(true)}}}

\newcommand{\stat}{\sin^22 \theta_{23}{\mbox {(true)}}}
\newcommand{\dcpt}{\delta_{CP}{\mbox {(true)}}}

\newcommand{\sig}{$3\sigma$}

\begin{document}

\thispagestyle{empty}
\begin{flushright}
\texttt{HRI-P-07-11-002}\\
\texttt{CU-PHYSICS-12/2007}\\
\end{flushright}
\bigskip

\begin{center}
{\Large \bf Unraveling neutrino parameters 
with a magical beta-beam experiment at INO} 

\vspace{.5in}

{\bf Sanjib Kumar Agarwalla$^{\star,\dagger,a}$, 
Sandhya Choubey$^{\star,b}$, Amitava Raychaudhuri$^{\star,\dagger,c}$}
\vskip .5cm
$^\star${\normalsize \it Harish-Chandra Research Institute,} \\
{\normalsize \it Chhatnag Road, Jhunsi, Allahabad  211019, India}\\
\vskip 0.4cm
$^\dagger${\normalsize \it Department of Physics, University of Calcutta,} \\ 
{\normalsize \it 92 Acharya Prafulla Chandra Road, Kolkata  700009, India}
\vskip 1cm
\vskip 2cm

{\bf ABSTRACT}
\end{center}
We expound in detail the physics reach of an experimental set-up 
in which the proposed large magnetized iron detector at the 
India-based Neutrino Observatory (INO) would serve as 
the far detector for a
so-called beta-beam. If this pure $\nue$ and/or 
$\anue$ beam is shot from some 
source location like CERN such that the source-detector 
distance $L \simeq 7500$ km, the impact of the CP phase 
$\delta_{CP}$ on the oscillation probability and associated 
parameter correlation and degeneracies are almost 
negligible. This ``magical'' beta-beam experiment would have 
unprecedented sensitivity to the neutrino mass hierarchy and 
$\theta_{13}$, two of the missing ingredients needed for 
our understanding of the neutrino sector. 
With Lorentz boost $\gamma=650$ and
irrespective of the true value of $\delta_{CP}$,
the neutrino mass hierarchy could be determined at $3\sigma$ C.L.
if $\sin^22\theta_{13}{\rm {(true)}} > 5.6 \times 10^{-4}$
and we can expect an unambiguous signal for $\theta_{13}$ 
at $3\sigma$ C.L. if $\sin^22\theta_{13}{\rm {(true)}} > 5.1 \times 10^{-4}$
independent of the true neutrino mass hierarchy.

\vskip 3cm

\noindent $^a$ email: sanjib@mri.ernet.in

\noindent $^b$ email: sandhya@mri.ernet.in 

\noindent $^c$ email: raychaud@mri.ernet.in

\newpage

\section{Introduction}

Neutrino physics has entered the precision era, with the thrust 
now shifting to detailed understanding of the structure of the 
neutrino mass matrix, accurate reconstruction of which 
would shed light on the underlying new physics that gives 
rise to neutrino mass and mixing. The full mass matrix is given 
in terms of nine parameters, the three neutrino masses, 
the three mixing angles and the three CP violating phases. 
Neutrino oscillation experiments are sensitive to only  
two mass squared differences, all the three mixing angles and 
the so-called Dirac CP violating phase. 
The remaining parameters,
comprising of the absolute neutrino mass scale and the 
two so-called Majorana phases, have to be determined 
elsewhere. 
We already have very good
knowledge on the two mass squared differences and two of the 
three mixing angles. Results from 
solar neutrino experiments \cite{solar}, 
which have been collecting  
data for more than four decades have now culminated in 
choosing the Large Mixing Angle (LMA) solution. The latest 
addition to this huge repertoire of experimental data is the 
result from the on-going Borexino experiment \cite{borex}, and this 
result is consistent with the LMA solution. 
This conclusion from solar neutrino experiments 
has been corroborated independently 
by the KamLAND reactor antineutrino experiment \cite{kl,kltalk}, 
and a combined analysis of the solar and KamLAND data 
gives as best-fit\footnote{We 
use a convention where $\Delta m_{ij}^2=m_i^2-m_j^2$.}  
$\ms=7.6\times 10^{-5}$ eV$^2$ and $\sss=0.32$ \cite{kltalk,limits}. 
The other mass squared difference $\ma$ and 
mixing angle $\theta_{23}$ are now pretty well determined 
by the zenith angle dependent atmospheric $\numu$ data in  
SuperKamiokande \cite{atm} and the long baseline experiments 
K2K \cite{k2k} and MINOS \cite{minos}. The combined data from 
the atmospheric and long baseline experiments have pinned 
down $|\ma| = 2.4\times 10^{-3}$ eV$^2$ and $\sin^22\theta_{23}=1$. 

Despite these spectacular achievements, 
a lot of information is still required 
to complete our understanding of the neutrino sector. 
While the solar neutrino data have 
confirmed that $\ms >0$ at a C.L. of more than $6\sigma$, we 
still do not know what is the sign of $\ma$. 
Knowing the ordering 
of the neutrino masses is of prime importance, because it dictates 
the structure of the neutrino mass matrix, and hence could 
give vital clues towards the underlying theory of neutrino masses 
and mixing. Knowing the $sgn(\ma)$ could have other far-reaching 
phenomenological consequences. For instance, if it turns out the 
$\ma < 0$ and yet neutrino-less double beta decay is not 
observed even in the very far future experiments, 
that would be a strong hint that the neutrinos 
are not Majorana particles (see for {\it e.g.} \cite{0vbbus} and 
references therein). 
Also, our knowledge on the third mixing angle 
$\theta_{13}$ is restricted to an upper bound of $\sch < 0.04$ 
from the global analysis of all solar, atmospheric, long baseline 
and reactor data, including the CHOOZ \cite{chooz} results in 
particular. 
Non-zero $\theta_{13}$ 
brings in the possibility of 
large Earth matter effects \cite{msw1,msw2,msw3}
for GeV energy accelerator 
neutrinos travelling over long distances. 
Effect of matter on neutrino oscillations 
depends on the \sgnma{} and is opposite for neutrinos and 
antineutrinos. For a given \sgnma{} it enhances 
the oscillation probability in 
one of the 
channels and suppresses it in 
the other. Therefore, comparing the 
neutrino signal against the antineutrino signal in very long 
baseline experiments gives us a powerful tool to 
determine \sgnma. 
A non-zero value of this mixing angle would also open up 
the possibility of detecting 
CP violation in the neutrino sector.

Tremendous effort is underway to determine 
$\theta_{13}$, the CP phase $\delta_{CP}$ and 
\sgnma{} using long baseline experiments 
\cite{t2k,nova,huber10,iss}. 
Future programs involving accelerator based neutrino beams 
include among others the T2K \cite{t2k}, NO$\nu$A \cite{nova},
Superbeams, beta-beams and Neutrino Factories \cite{iss}.  
In earlier papers \cite{paper1,betaino} 
we have proposed and expounded 
the possibility of measuring to a very high degree of 
accuracy the mixing angle $\theta_{13}$ and \sgnma{} {\it aka},
the neutrino mass hierarchy\footnote{What we usually refer to as 
the neutrino mass hierarchy is really the neutrino mass ordering. 
Therefore, our discussion and results are equally relevant for a
quasi-degenerate neutrino mass spectrum as they are for hierarchical 
and inverted hierarchical spectra.}, 
in an experimental set-up 
where a pure and intense $\nue$ and/or $\anue$ beam is shot 
from CERN to the India-based Neutrino Observatory (INO) \cite{ino}. 
This pure and intense source of $\nue$ and/or $\anue$ 
flux could be the so-called beta-beam \cite{zucc}, 
which is created when fully 
ionized and highly accelerated radioactive ions 
beta decay in the straight sections of a ring, where they 
are circulated and stored, after being produced, 
collected, bunched and accelerated.  
A large magnetized iron calorimeter (ICAL)
is expected to come-up soon at  
the INO facility in India. Since the energy threshold of this 
detector would be at least 1 GeV, the beta-beam 
should necessarily be a multi-GeV beam. While the most 
widely discussed source  
ions for beta-beams, $^{18}Ne$ and $^6He$, need very large 
values of the Lorentz boost to reach 
multi-GeV energies \cite{paper1},
alternative ions with larger end-point energy, such as 
$^8B$ and $^8Li$, can be used with reasonable acceleration.
While the most discussed design 
for the beta-beam set-up \cite{cernmemphys}
needs modest Lorentz boosts, it suffers from 
the effect of the so-called ``parameter degeneracies'' 
\cite{intrinsic,minadeg,th23octant} giving rise to 
eight-fold degenerate ``clone solutions'' \cite{eight}.
A very big advantage that the CERN-INO beta-beam experiment
would have is that at the baseline of 7152 km,
the $\delta_{CP}$ dependent terms (almost) drop out 
from the expression for the $\nue\rightarrow \numu$ 
oscillation probability. 
As a result, this experiment is free of 
two of the three 
degeneracies \cite{eight,magic,magic2}. 
In addition, large energies and the large distance involved 
allows the neutrino to pick up near-resonant matter effects, 
enhancing the oscillation probability which can thereby
compensate to a great extent the reduction of the 
neutrino flux due to the $1/L^2$ factor \cite{betaino}. 
This makes the CERN-INO beta-beam set-up 
almost magical for determining $sgn(\ma)$ and $\theta_{13}$. 

In \cite{betaino} we studied the physics potential 
of this experimental set-up when we run the beta-beam in 
{\it only one polarity} for five or ten years. That is, 
we probed the sensitivity of the experiment to 
$sgn(\ma)$ and $\theta_{13}$ using {\it only} the $\nue$ ($\anue$)
beam for five years running with $1.1\times 10^{18}$ 
($2.9\times 10^{18}$) useful ion decays per year.
In this paper we extend our analysis by including data from 
{\it both} the neutrino and antineutrino run of this beta-beam 
experimental set-up. We demonstrate how adding data from both 
polarities 
serves to strengthen precisely those regions where the individual 
ones are less powerful and thus 
significantly enhances the mass hierarchy sensitivity 
of the experiment. As a further refinement, 
we analyse the full spectral data expected in the CERN-INO 
beta-beam set-up. We present and compare results from the 
rates-only analysis against results from the analysis where 
we incorporate the full spectrum. 
For $\theta_{13}$ measurement, we 
consider two scenarios: (i) when there are no 
$\numu$ (or $\anumu$) events in the detector and (ii) 
when we see a signal in the detector. In the former case 
we present the $3\sigma$ upper bound on $\stch$
expected from the null results. We call 
this the ``$\stch$ sensitivity reach'' of the experiment. 
In the latter case, we first study the range of 
``true'' values\footnote{Throughout this paper
we denote true value of the parameters by putting ``(true)'' 
in front of them.} of $\stcht$ 
for which the experiment would be
able claim to have seen a signal at the $3\sigma$ C.L.
This is termed as the ``$\stcht$ discovery reach'' of the 
experiment. Finally, for a non-zero signal at the detector, 
we study how precisely $\stch$ can be determined with 5 years 
of combined neutrino and antineutrino run. 
In our earlier paper \cite{betaino}, 
we had given all results assuming that $\delta_{CP}{\rm(true)}=0$.
The hierarchy and $\theta_{13}$ measurement sensitivities 
however depend on the value of  $\delta_{CP}{\rm(true)}$.
In this paper we find the physics reach of the CERN-INO 
beta-beam set-up for  
all possible values of $\delta_{CP}{\rm(true)}=0$
and show the best and worst possible physics reach.
We also study the impact of changing the number of useful 
ion decays in the straight sections of the beta-beam storage 
ring and compare the dependence of the sensitivity on 
the number of ion decays and the Lorentz boost $\gamma$. 

For all results presented in this paper, we use the full 
PREM Earth matter density profile \cite{prem} for simulating 
the prospective data. 
When we fit this simulated data, we 
allow for a 5\% uncertainty in the PREM profile and  
take it into account by inserting a prior and 
marginalizing over the density 
normalization. We also study the impact of changing the 
Earth matter density by $\pm 5$\% in the data itself.
We also study the impact of changing the energy threshold of 
the detector and the background rejection factor. 
In our analysis here 
we also allow the parameters $\ms$ and $\sss$
to vary arbitrarily in the fit. 

We begin by providing a brief overview of 
the proposed experimental set-up, the expected event rate, and 
the importance of the magic baseline in section 2. 
In section 3, we give our results on the sensitivity of the 
CERN-INO beta-beam set-up to the neutrino mass hierarchy.
In section 4, we discuss the potential of measuring/constraining 
$\theta_{13}$. We end with our conclusions in section 5. The details 
of our numerical code and analysis procedure are relegated to an
Appendix.

\section{The CERN-INO beta-beam Experimental Set-up}

\begin{figure}[t]
\includegraphics[width=8.0cm, height=7.0cm]{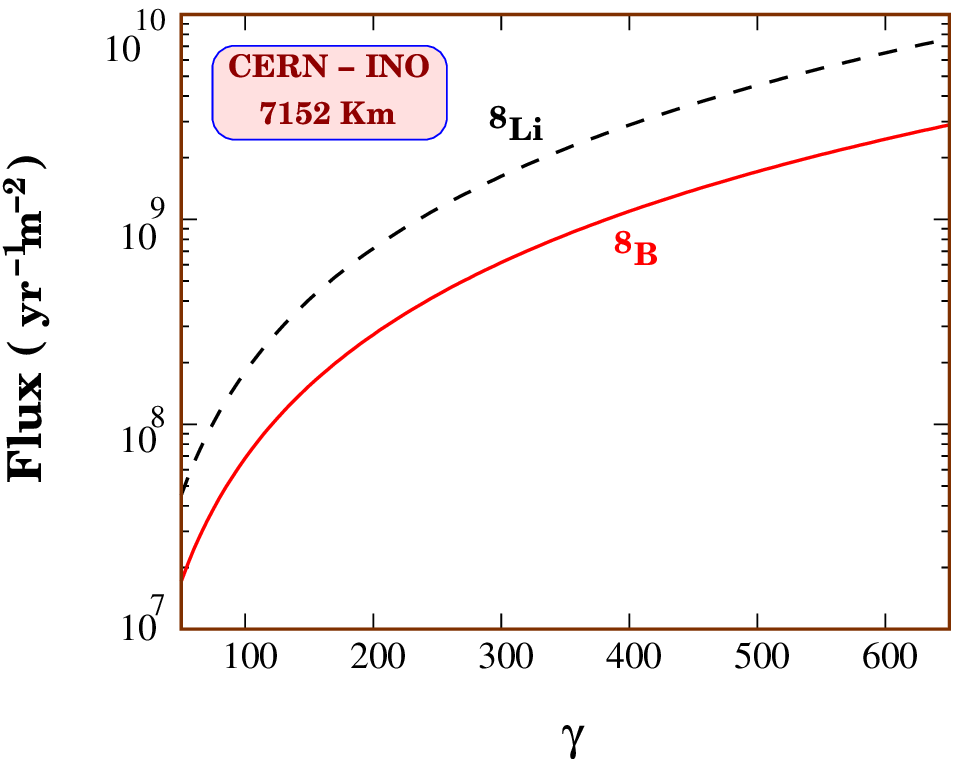}
\vglue -7.0cm \hglue 8.8cm
\includegraphics[width=8.0cm, height=7.0cm]{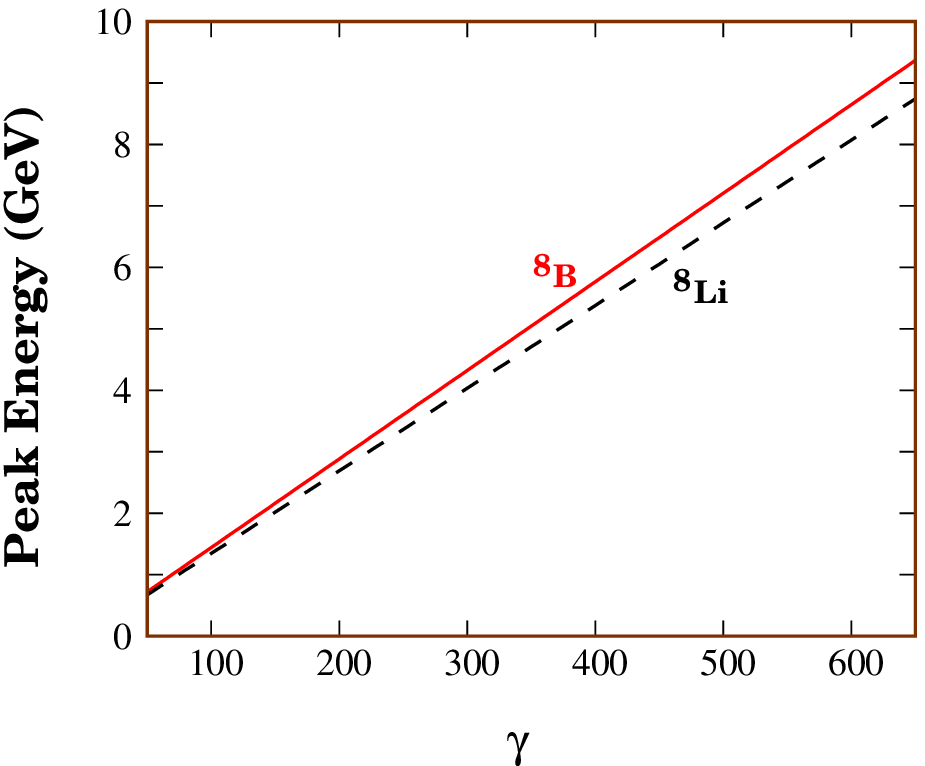}
\caption{\label{fig:flux}
Left panel shows 
the total flux in $yr^{-1}m^{-2}$ expected at INO, as a function of 
the Lorentz factor $\gamma$. The solid (dashed) 
line corresponds to $^8B$ ($^8Li$) and we have assumed 
$1.1\times 10^{18}$ ($2.9\times 10^{18}$) useful ions 
decays per year. Right panel shows the energy at which the 
flux peaks, as a function of $\gamma$. 
}
\end{figure}
%

\begin{figure}[t]
\begin{center}
\includegraphics[width=10.0cm]{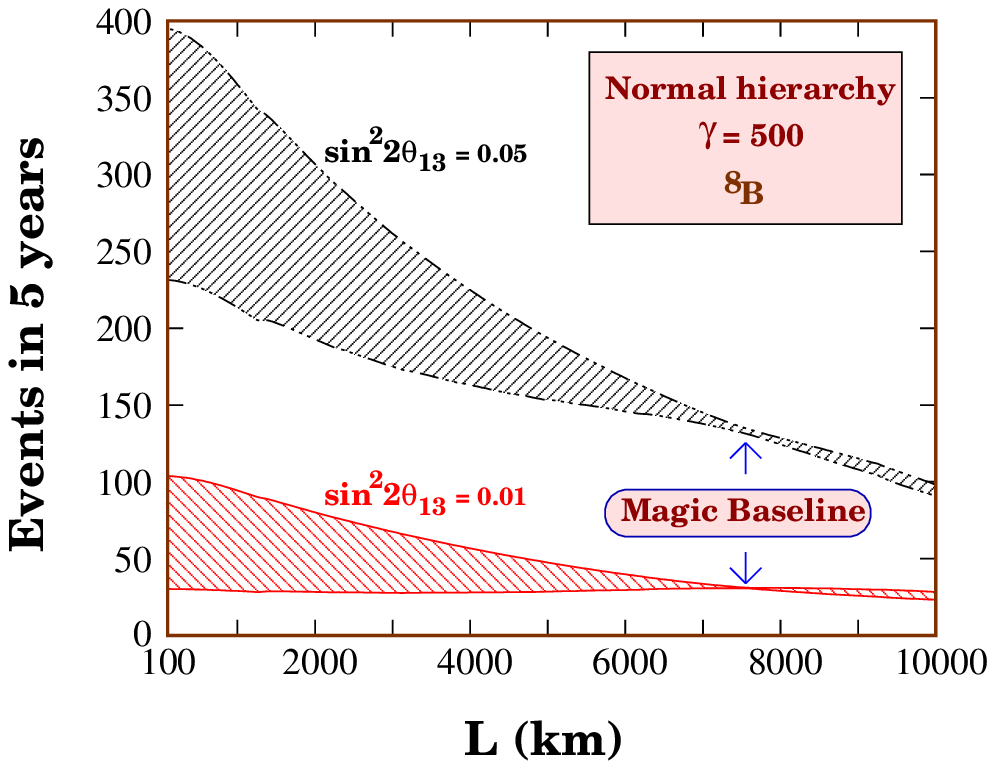}
\caption{\label{fig:magic}
Total number of expected 
events in five years as a function of the 
baseline $L$ for the $^8B$ source with $\gamma=500$ and for 
two values of $\stch$ and assuming that the 
normal hierarchy is true. 
The hatched areas show the 
expected uncertainty due to the CP phase. 
}
\end{center}
\end{figure}
%

\begin{figure}[t]
\begin{center}
\includegraphics[width=15.0cm]{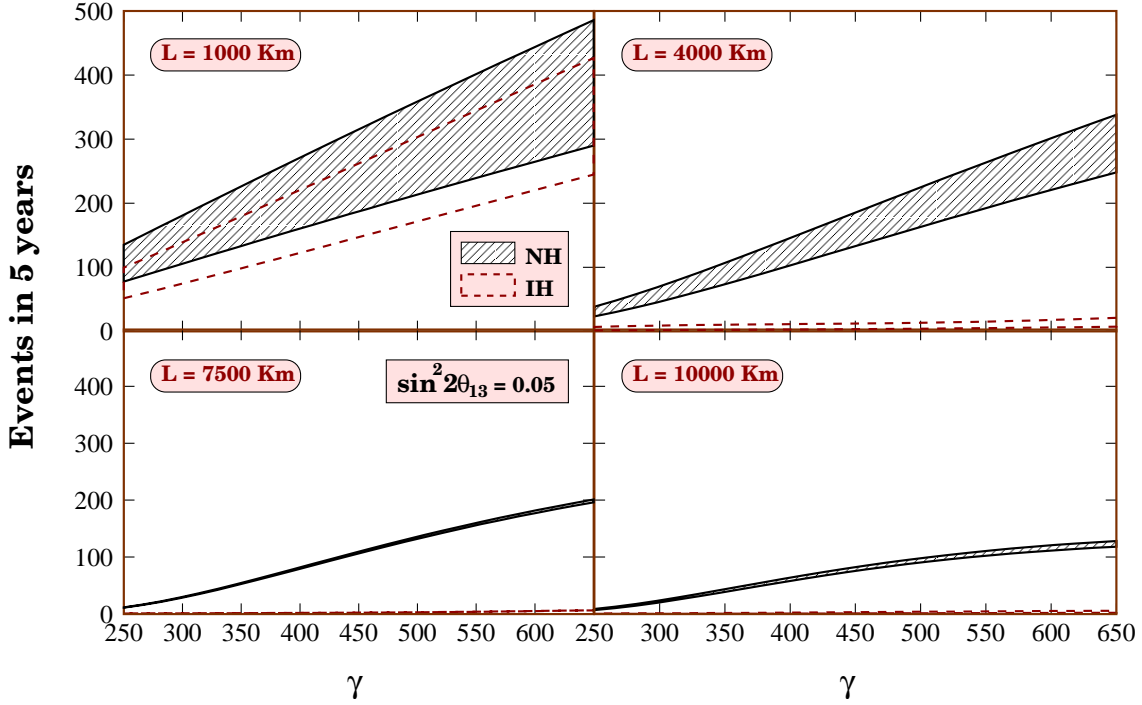}
\caption{\label{fig:eventshier}
Total number of events as a function of $\gamma$ for the 
$^8B$ source, for different values of $L$ are shown in 
the four panels. The black hatched area shows the uncertainty 
range due to the CP phase when NH is true, 
while the area between the maroon dashed lines 
shows the corresponding 
uncertainty when IH is true.
For all cases we assume $\stch=0.05$.
}
\end{center}
\end{figure}
%

\begin{figure}[t]
\begin{center}
\includegraphics[width=15.0cm]{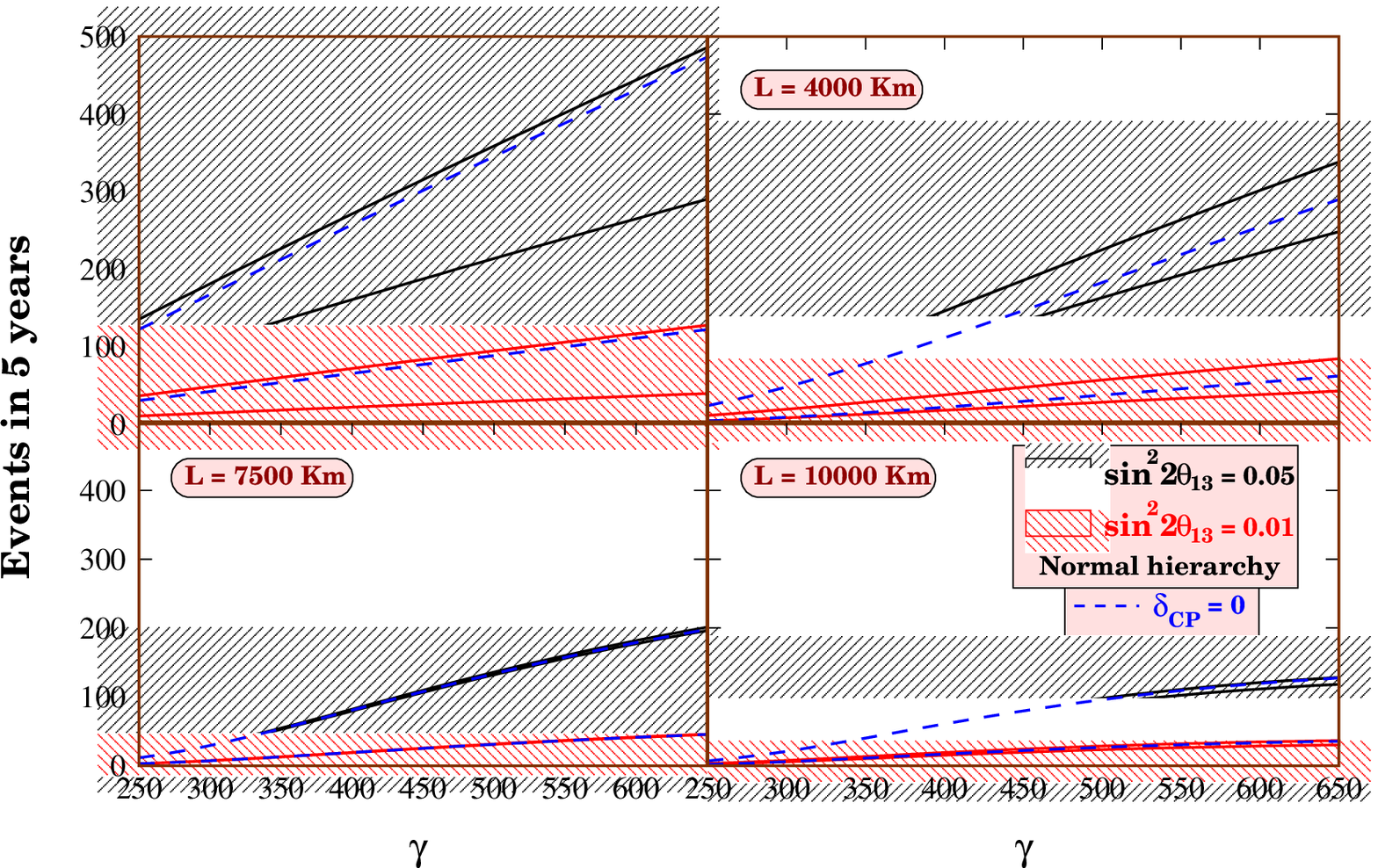}
\caption{\label{fig:eventsth13}
Total number of events as a function of $\gamma$ for the 
$^8B$ source, for different values of $L$ are shown in 
the four panels. 
The black hatched area shows the uncertainty 
range in the events due to CP phase when $\stch=0.05$, 
while the 
red hatched area 
shows the corresponding 
uncertainty when $\stch=0.01$.
For all cases we assume NH to be true.
}
\end{center}
\end{figure}
%

\begin{figure}[t]
\begin{center}
\includegraphics[width=10.0cm]{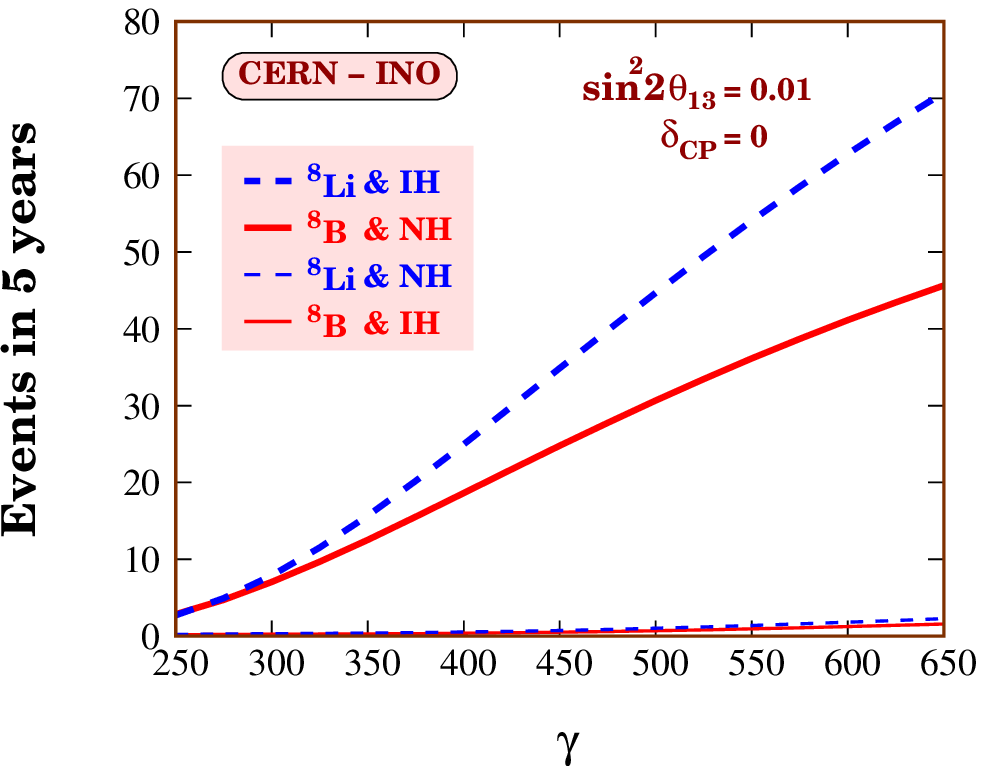}
\caption{\label{fig:eventsbetaino}
Total number of events as a function of $\gamma$ for the 
$^8B$ (solid lines) and 
the $^8Li$ (dashed lines) 
sources. Results for both normal and inverted hierarchies are shown.
}
\end{center}
\end{figure}
%

Very pure and intense $\nue$ and/or $\anue$ beams can be 
produced by the decay of highly accelerated 
radioactive beta unstable ions, circulating in a 
storage ring. This is what is called a ``beta-beam'' and 
was first proposed by Piero Zucchelli \cite{zucc}. 
Since the flux spectrum is determined entirely by the end-point 
energy of the parent ions and the Lorentz boost provided 
by the acceleration, it is almost free of 
systematic uncertainties. The flux normalization is 
determined by the number of useful ion decays in the 
straight section of the beam.  
The selection of the beta unstable parent ion is determined 
by a variety of factors essential for efficiently producing, 
bunching, accelerating and storing these ions in the 
storage ring. Among the candidate ions discussed in the 
literature, the ones which have received most attention 
so far are the $^{18}Ne$ and $^6He$ ions for producing 
the $\nue$ and $\anue$ beams respectively 
\cite{cernmemphys,betaoptim,bc,volpe,oldpapers,doninibeta}. 
Both these 
ions have a similar end-point energy which is about 
4 MeV in the parent ion's rest frame. Two other 
candidates, $^8B$ and $^8Li$, as source ions for 
$\nue$ and $\anue$ beams respectively,  
have been recently shown to be viable \cite{rubbia,mori,doninialter,rparity,olga}. 
The advantage that these ions have over $^{18}Ne$ 
and $^6He$ is their larger end-point energies, which is 
higher by a factor of more than 3. 
Therefore, for the 
same Lorentz boost factor, we expect the resultant 
$^8B$ and $^8Li$ 
beams to be about 3 times higher in energy compared to 
the $^{18}Ne$ and $^6He$ beams.  
We refer the reader to 
Table 1 of our earlier paper \cite{betaino} for the full 
details about 
characteristics of the four beta-beam candidate ions.

These large beta decay end-point energy ions are particularly 
important for the CERN-INO beta-beam experimental set-up 
discussed in \cite{paper1,betaino}, where the beta-beam 
from CERN is shot to a magnetized iron detector 
at the India-based neutrino observatory (INO). 
The INO will be located in southern India, close to the 
city of Bangalore. The CERN to INO 
distance corresponds to 7152 km, which is tantalizingly close 
to the ``magic baseline'' \cite{magic,magic2} (see also \cite{petcov}). 
Being free of the problem of parameter degeneracies
\cite{intrinsic,minadeg,th23octant,eight}, 
the magic baseline is known to be particularly useful for 
measuring the mixing angle $\theta_{13}$ and
$sgn(\ma)$. The concept of the magic baseline can be 
very easily understood by looking at the approximate 
expression of $P_{e\mu}$, where the conversion 
probability is expanded in the small parameters 
$\theta_{13}$ and $\alpha \equiv \ms/\ma$, keeping 
only terms up to second order in these small parameters. 
This expression for the ``golden channel''  \cite{golden}
probability is given as \cite{golden,freund}
\be
 P_{e\mu} &\simeq& 
 \sin^2\theta_{23} \sin^22\theta_{13}
\frac{\sin^2[(1-\hat{A})\Delta]}{(1-\hat{A})^2}\nonumber \\
&+& \alpha \sin2\theta_{13} \sin2\theta_{12} \sin2\theta_{23} 
\sin\delta_{CP} \sin(\Delta) \frac{\sin(\hat{A}\Delta)}{\hat{A}}
\frac{\sin[(1-\hat{A})\Delta]}{(1-\hat{A})} \nonumber \\
&+& \alpha \sin2\theta_{13} \sin2\theta_{12} \sin2\theta_{23} 
\cos\delta_{CP} \cos(\Delta) \frac{\sin(\hat{A}\Delta)}{\hat{A}}
\frac{\sin[(1-\hat{A})\Delta]}{(1-\hat{A})} \nonumber \\
&+& \alpha^2 \cos^2\theta_{23} \sin^22\theta_{12} 
\frac{\sin^2(\hat{A}\Delta)}{{\hat{A}}^2}
,
\label{eq:pemu}
\ee
where 
\be
\Delta\equiv \frac{\ma L}{4E},
~~
\hat{A} \equiv \frac{A}{\ma},
\label{eq:matt}
\ee
where $A=\pm 2\sqrt{2}G_FN_eE$ is the matter potential, $N_e$ being the 
electron number density inside the earth and $G_F$ the 
Fermi constant, the $+$ sign 
refers to neutrinos while the $-$ to antineutrinos. 
In Eq. (\ref{eq:pemu}) the second term has the CP violating 
part. The CP phase $\delta_{CP}$ is positive for neutrinos and 
negative for antineutrinos and therefore the $\sin\delta_{CP}$ 
term changes sign. The third term, though $\delta_{CP}$ 
dependent, is CP conserving, while the fourth term is independent 
of $\theta_{13}$ as well as $\delta_{CP}$. 
If there exists a baseline $L$ for which the condition 
\be
\sin(\hat{A}\Delta)=0
\label{eq:condmagic}
\ee
holds, then the second, third and fourth 
terms in Eq. (\ref{eq:pemu}) drop out and one is left with 
just the first term. In particular, we see that all 
$\delta_{CP}$ dependent terms go away at this magic baseline.
Therefore, this experimental set-up is free of two of the 
three parameter degeneracies, providing us with a firm 
bedrock for determining $\theta_{13}$ and the mass hierarchy
\cite{magic,betaino}. It turns out that for the PREM 
Earth matter density profile, 
the condition given by Eq. 
(\ref{eq:condmagic}) is satisfied for $L\simeq 7500$ km.  
In \cite{betaino} we also stressed the 
point that for this very long baseline, 
neutrinos would also pick up large, and 
possibly near-resonant, matter effects. Largest 
enhancement of oscillations due to matter effects 
of course comes about when the product of the 
mixing angle term in matter and the $(\ma)^M$ 
driven oscillatory term is the largest \cite{gandhi,pee}.
For the CERN-INO baseline of 7152 km, the probability 
for largest conversion is expected for $E\simeq 6$ GeV, for 
$\stch=0.01$ and $\ma=2.5\times 10^{-3}$ eV$^2$.

The ICAL detector, which will be a 50 kton magnetized 
iron calorimeter, will be built at INO \cite{ino}.
There is a possibility that the detector mass 
might be increased to 100 kton at a later stage. 
The approved design of the detector comprises of 
6 cm iron slabs interleaved with 2 cm thick 
glass Resistive Plate Chambers (RPC), which would serve 
as the active detector elements for ICAL. 
The iron will be magnetized by an external field 
of about 1 Tesla, giving the detector charge identification 
capability. The detection efficiency of ICAL 
after cuts is expected to be about 80\% and energy threshold 
would be about 1 GeV. 
In what follows, we will use an energy threshold of 1.5 GeV for 
our main results. However, we will also show the impact of 
changing the threshold. The energy resolution of the detector 
is expected to be reasonable and we assume that the 
neutrino energy will be reconstructed with an uncertainty parameterized 
by a Gaussian energy resolution function with a HWHM $\sigma_E=0.15E$, 
where $E$ is the energy of the neutrino. We will present
and compare the sensitivity of this experimental set-up 
with and without the full spectral analysis. Details of 
our numerical approach can be found in \cite{betaino} and 
in the Appendix.

In the left panel of 
Fig. \ref{fig:flux} we show the energy integrated 
total number of 
neutrinos in units of $yr^{-1}m^{-2}$ 
arriving at INO, as a function of the Lorentz boost 
$\gamma$. 
The solid (dashed) 
line corresponds to $^8B$ ($^8Li$) and we have assumed 
$1.1\times 10^{18}$ ($2.9\times 10^{18}$) useful ion
decays per year\footnote{Unless stated otherwise these are the
reference luminosities in all the figures. Also, all figures
correspond to a five year run.}. Throughout this paper we have assumed 
that the detector is aligned along the axis of the beam.
The figure shows that the energy integrated flux 
arriving at the detector increases almost quadratically with 
$\gamma$. Note that with the same 
accelerator, the Lorentz boost acquired by $^8B$ is 
1.67 times larger than that by $^8Li$, determined 
by the charge to mass ratios of the ions.
The right panel of Fig. \ref{fig:flux} 
depicts the energy at which the flux peaks,
as a function of the Lorentz boost 
$\gamma$. It turns out that this peak energy is roughly 
half the 
maximum energy of 
the beam, which is given as $E_{\rm max} \simeq 2(E_0-m_e)\gamma$.

In Fig. \ref{fig:magic} we show the number of events expected 
in five years as a function of the baseline $L$, 
if we run the experiment in the neutrino mode with 
$\gamma=500$. A similar figure is expected for the antineutrino
beam. The upper black hatched area shows the 
events for $\stch=0.05$ and 
the lower red hatched area corresponds to $\stch=0.01$. 
For each baseline $L$, the range  
covered by the hatched area shows the uncertainty in the 
expected value of the number of events due to the 
completely unknown $\delta_{CP}$, which could 
take any value from 0 to 2$\pi$. The baseline $L$ where 
the width of this band reduces to (almost) zero is 
the magic baseline. We see from the figure that the magic 
baseline appears at about $L\simeq 7500$ km. Note that while 
for $\stch=0.01$ the magic baseline is very clearly defined 
with the CP dependence going completely to zero, 
for the higher value of $\stch$ of 0.05, the 
``magic'' is not complete. 
The reason for this anomaly can be traced to the fact that 
Eq. (\ref{eq:pemu}) was derived for only very small values of 
$\theta_{13}$. For larger values of this angle, 
higher order terms become important.  These terms might 
depend on $\delta_{CP}$ and remain 
non-zero even at the magic baseline. 

In Figs. \ref{fig:eventshier} and \ref{fig:eventsth13}
we show the impact of the magic baseline on the mass 
hierarchy and $\theta_{13}$ sensitivity respectively. 
In each of the four panels of both the figures we show the 
expected events in five years as a function of $\gamma$.
Each panel is for a certain fixed value of $L$, shown 
in the corresponding panel. In Fig. \ref{fig:eventshier}
the black hatched area shows the band for normal hierarchy 
(NH) while the open band delimited by the dashed red lines 
are for the inverted hierarchy (IH). As in Fig. \ref{fig:magic}
the band correspond to the uncertainty in the event rate 
due to the unknown $\delta_{CP}$. The effect of the 
uncertainty of $\delta_{CP}$ almost vanishes for $L=7500$ km 
which is very close to the magic baseline. 
We can see that for the smaller baseline $L=1000$ km, NH and IH 
predictions are largely overlapping, making it almost 
impossible for these experiments to give sensitivity 
to the mass hierarchy unless $\stch$ turns out to be 
extremely large and $\delta_{CP}$ favorable. 
The hierarchy sensitivity is expected to improve as we
go to larger baselines and this is reflected from the 
two bands for NH and IH separating out. It turns out that 
because the matter effects are very large for the magic 
baseline and effect of CP uncertainty is zero, this baseline 
gives the best sensitivity to the mass hierarchy. For $L$ 
larger than magic, matter effects are higher but the 
flux is lower, while for $L$  lower than magic, 
flux is higher but the matter effects are lower. For 
both above and below the magic baseline, the effect 
of $\delta_{CP}$ is expected to further reduce the 
sensitivity. This is particularly true for the lower $L$ 
baselines. 
Fig. \ref{fig:eventsth13} shows the bands for NH but with two
different choices for $\stch$. Here the effect of the magic 
baseline is seen even more clearly. 

In Fig. \ref{fig:eventsbetaino} we show 
as a function of $\gamma$, 
the number of events 
expected in five years in the CERN-INO beta-beam set-up. 
The solid (dashed) 
lines are for neutrino 
(antineutrino) events, with the thick line 
showing the event rate for NH (IH) while the thin
line is for the IH (NH). We have assumed $\stch=0.01$ and 
$\delta_{CP}=0$. 
One point which is transparent from this figure and which will 
be very relevant in understanding the behavior of the CERN-INO 
beta-beam set-up is the following: For a given value of $\theta_{13}$ 
and for NH (IH), we expect a large 
number of events in the neutrino (antineutrino)
channel and almost negligible 
events in the antineutrino (neutrino) channel. 
This means that for NH (IH) it will be
the neutrino (antineutrino) channel which will 
be statistically more important. 

As discussed in detail in \cite{betaino} we expect hardly 
any background events in the CERN-INO beta-beam experiment. 
The atmospheric neutrino flux falls steeply with energy 
and is expected to produce much fewer events 
for the energy range that we are interested in\footnote{We 
will show in the next section that even a threshold energy 
of 4 GeV is easily 
admissible in our set-up, and above 
4 GeV there are much fewer atmospheric events.}. The fact that 
INO has charge identification capability further reduces the 
atmospheric background. The most important handle on the 
reduction of this background comes from the timing information 
of the ion bunches inside the storage ring. 
For 5T magnetic field and $\gamma=650$ for $^8B$ ions, the 
total length of the storage ring turns out to be 
19564 m. We have checked that with 
eight bunches inside this ring at any given time, a   
bunch size of about 40 ns would give an  
atmospheric background to {\it signal} ratio of about $10^{-2}$,
even for a very low $\stch$ of $10^{-3}$. For a smaller 
bunch span, this will go down even further. In addition, 
atmospheric neutrinos will be measured in INO during deadtime and 
this can also be used to subtract them out. 
Hence we do not include this negligible background here. 
In our numerical analysis we have assumed that the background 
events come only from the neutral current interactions of the 
beta-beam in the detector. We estimate it 
by assuming an energy independent  
background fraction of $\sim 10^{-4}$ \cite{mind}. 
We have noted that after five years of running of the CERN-INO 
beta-beam experiment with $\gamma=650$, we expect only 
about 0.1 background events. Nevertheless we take  this 
background into account in our numerical analysis.
Details of our numerical analysis are given in the Appendix.

\section{Measurement of the Neutrino Mass Hierarchy}

\begin{figure}[t]
\includegraphics[width=8.0cm, height=7.0cm, angle=0]{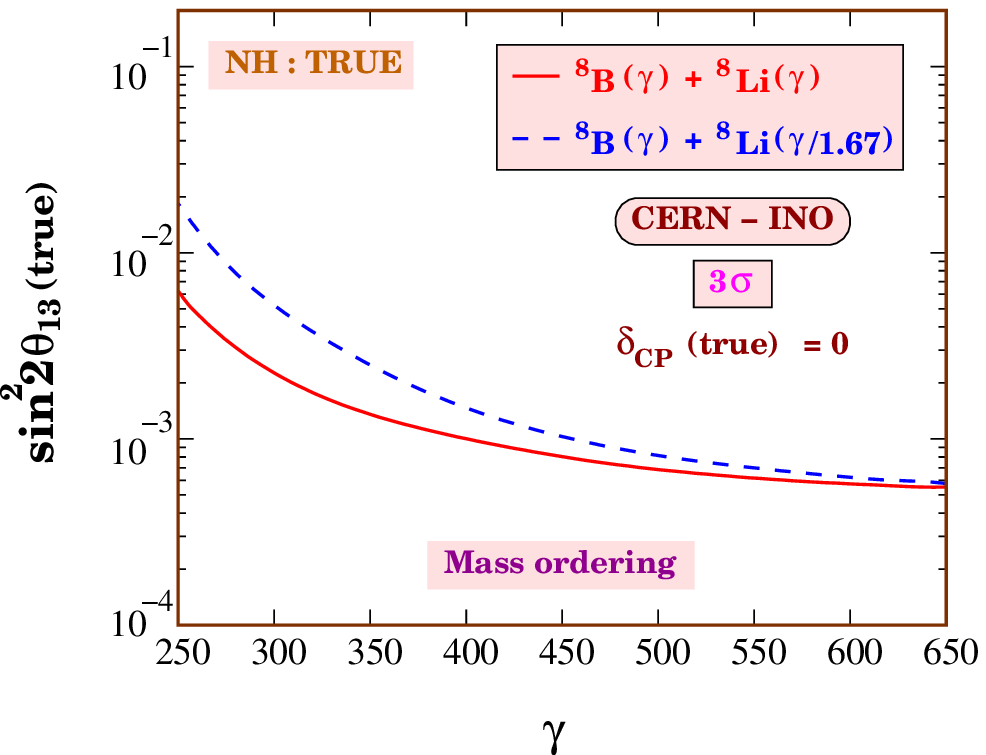}
\vglue -7.0cm \hglue 8.8cm
\includegraphics[width=8.0cm, height=7.0cm, angle=0]{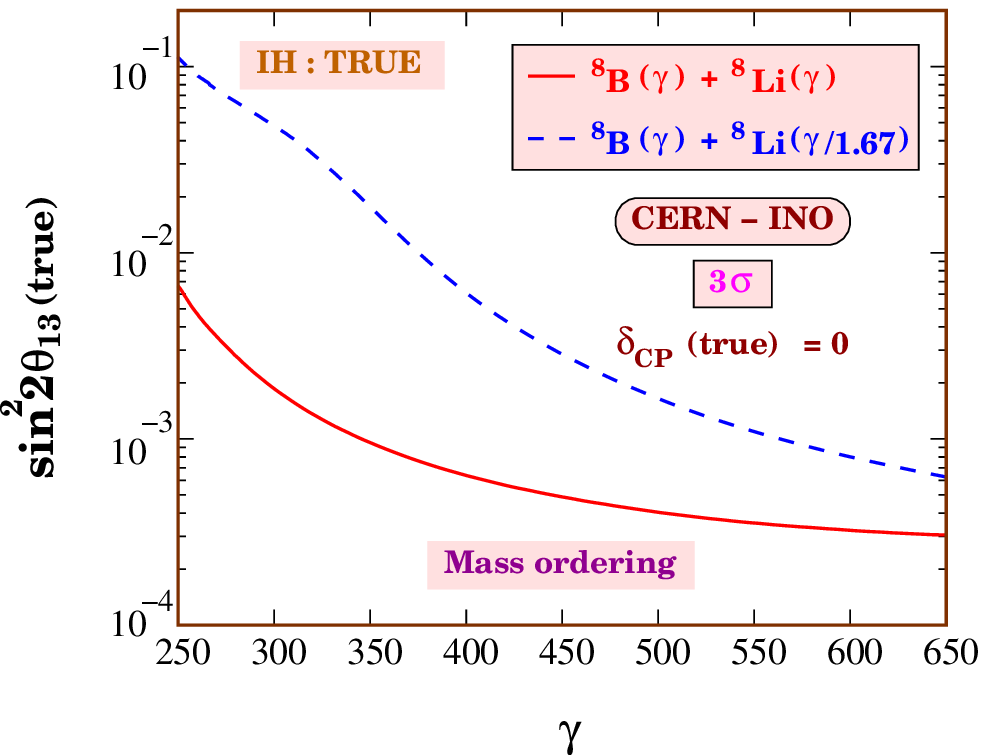}
\caption{\label{fig:senshier}
Minimum value of $\stch$(true) for which the wrong 
hierarchy can be ruled out at the $3\sigma$ C.L., 
as a function of 
$\gamma$. The left panel is for normal hierarchy as true, while 
the right panel is when inverted hierarchy is true. The 
red solid curves
show the sensitivity when the $\gamma$ is chosen to be the same 
for both the neutrino and the antineutrino beams. The 
blue dashed lines 
show the corresponding sensitivity when the $\gamma$ for the 
antineutrinos is scaled down by a factor of 1.67 with respect to the 
$\gamma$ of the neutrino beam.
}
\end{figure}
%

\begin{figure}[p]
\includegraphics[width=8.0cm, height=7.0cm]{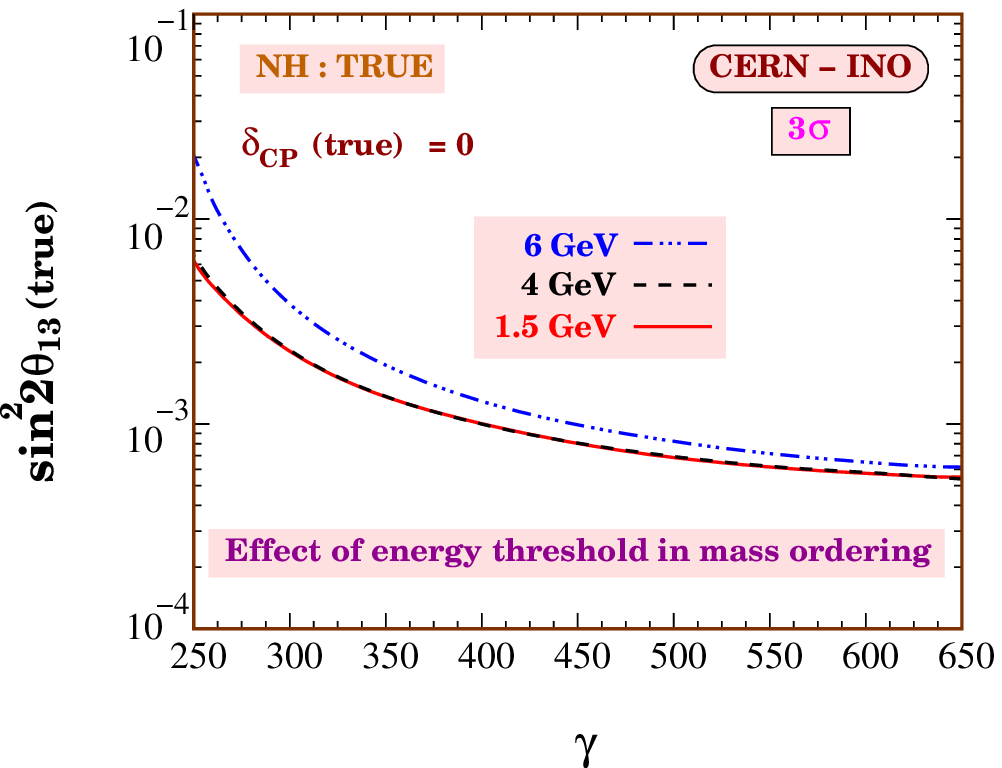}
\vglue -7.0cm \hglue 8.8cm
\includegraphics[width=8.0cm, height=7.0cm]{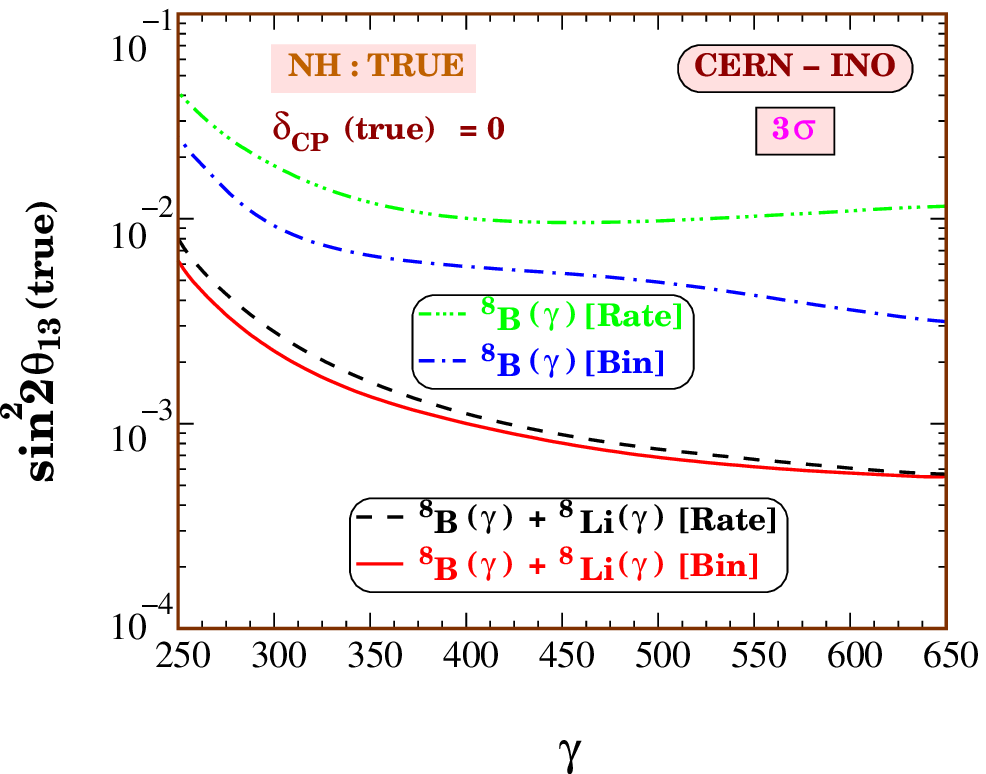}
\vglue -0.0cm \hglue 0.0cm
\includegraphics[width=8.0cm, height=7.0cm]{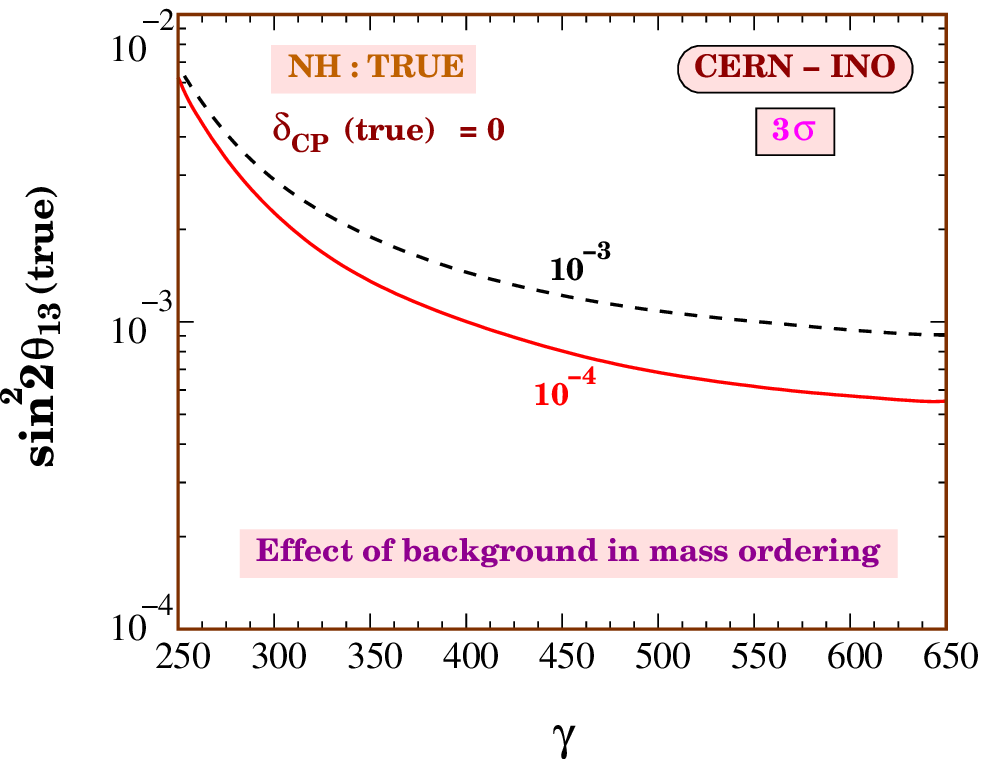}
\vglue -7.0cm \hglue 8.8cm
\includegraphics[width=8.0cm, height=7.0cm]{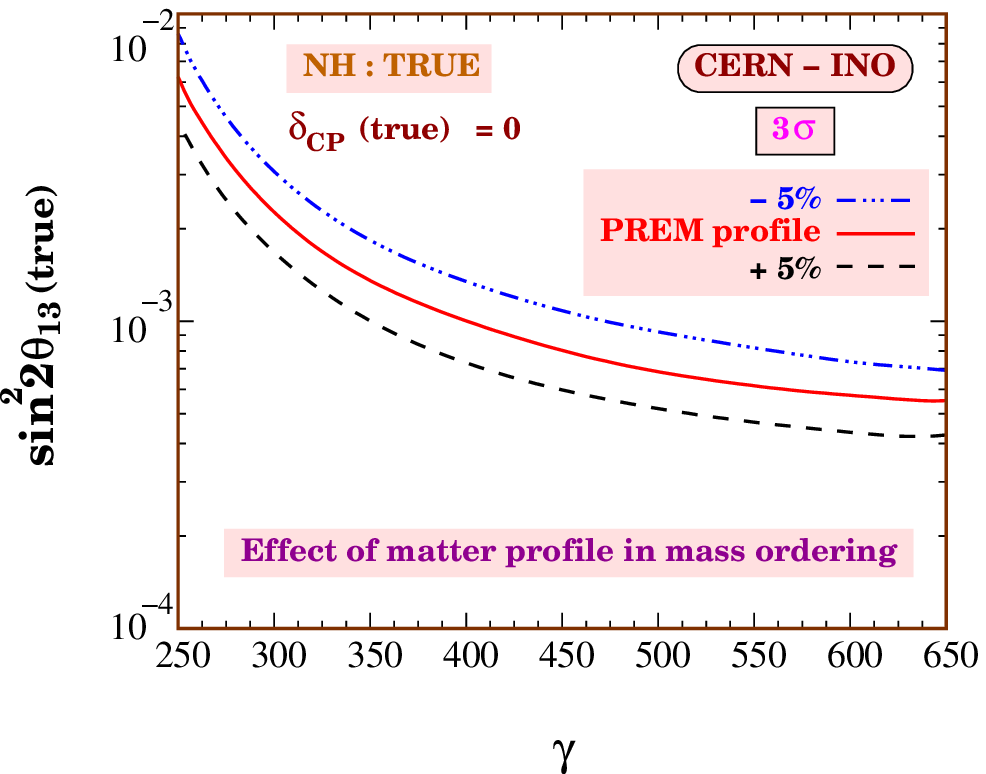}
\caption{\label{fig:senshiertest}
Plots showing the impact of various factors on the mass hierarchy 
sensitivity of the CERN-INO beta-beam experiment. The top left panel 
shows the impact of changing the detector threshold. The lower
left panel shows the effect of changing the background rejection 
factor. The top right panel shows the difference in the sensitivity 
between the rate and spectral analysis. The lower right panel 
shows how the density profile would impact the hierarchy sensitivity.
}
\end{figure}
%

\begin{figure}[t]
\includegraphics[width=8.0cm, height=7.0cm]{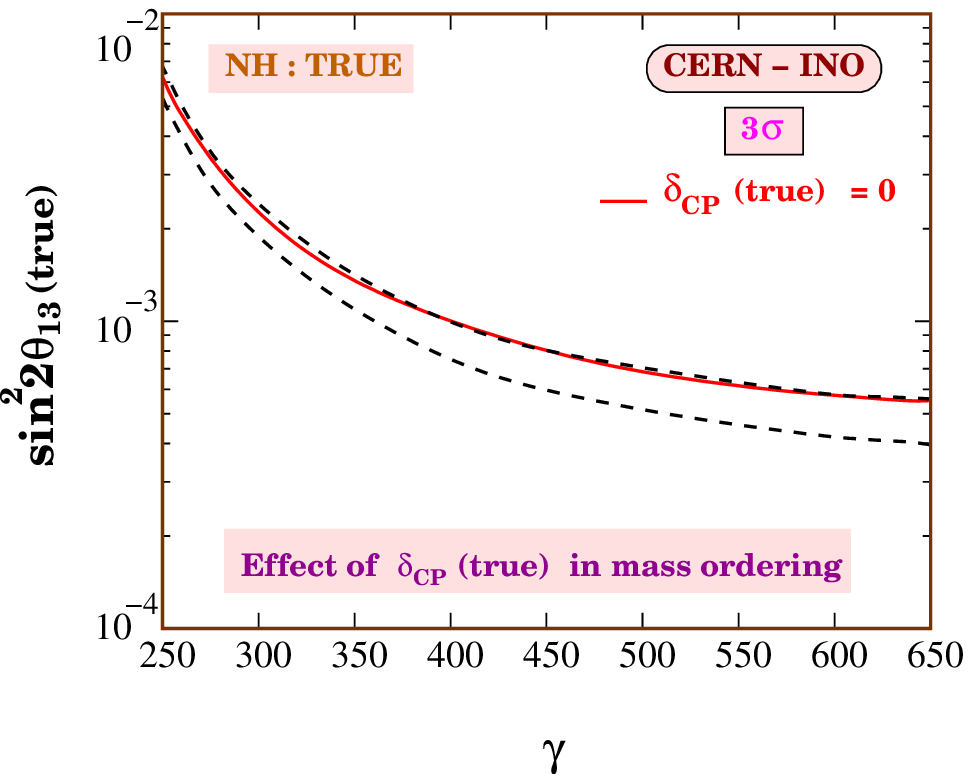}
\vglue -7.0cm \hglue 8.8cm
\includegraphics[width=8.0cm, height=7.0cm]{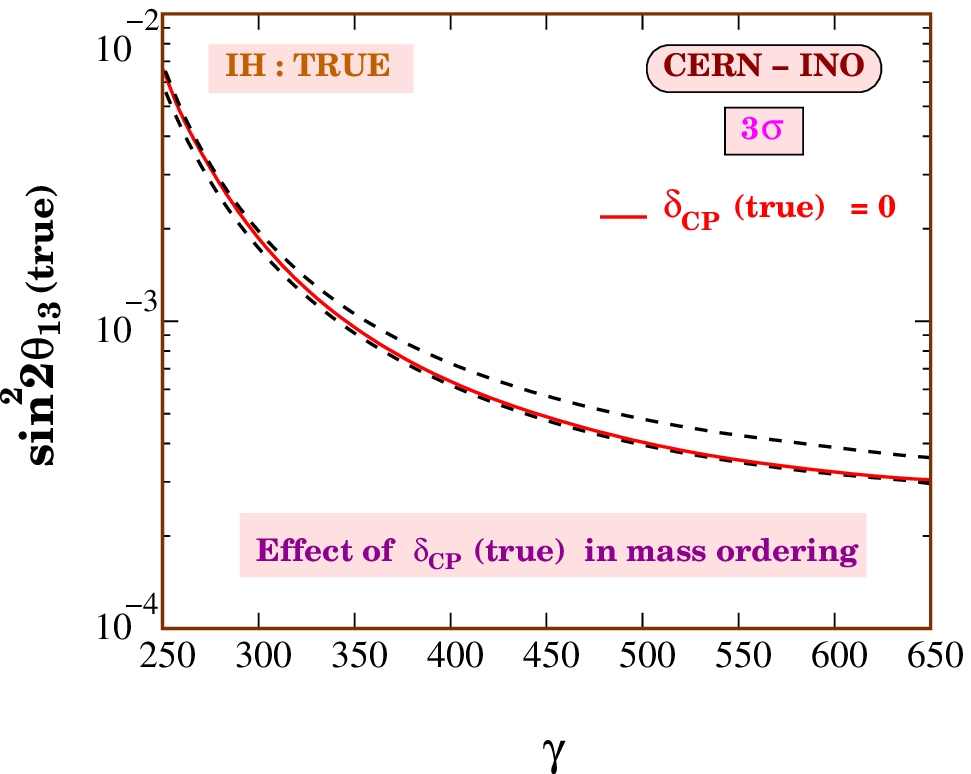}
\caption{\label{fig:senshiercploop}
Effect of $\dcpt$ on the hierarchy sensitivity. 
The black dashed
lines show the worst and best 
cases when we allow $\dcpt$ to take any 
value between 0 and 2$\pi$. The red solid curve corresponds 
to the reference case where $\dcpt=0$.
The left panel shows the 
case for true normal hierarchy while the right panel is for 
true inverted hierarchy.  
}
\end{figure}
%

\begin{figure}[t]
\includegraphics[width=8.0cm, height=7.0cm]{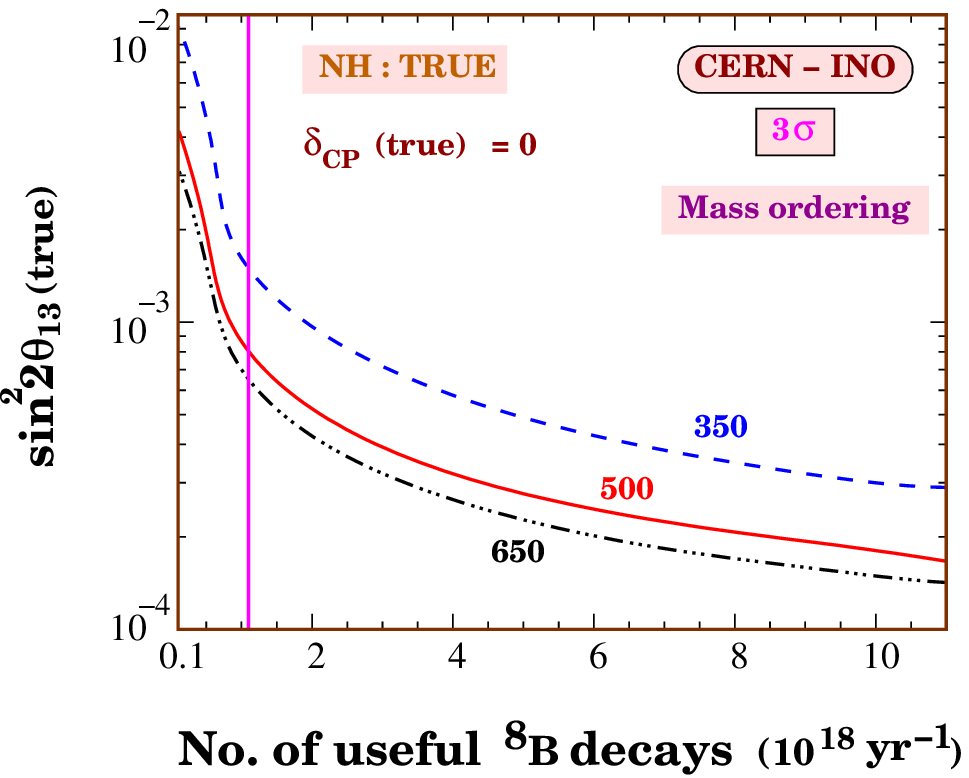}
\vglue -7.0cm \hglue 8.8cm
\includegraphics[width=8.0cm, height=7.0cm]{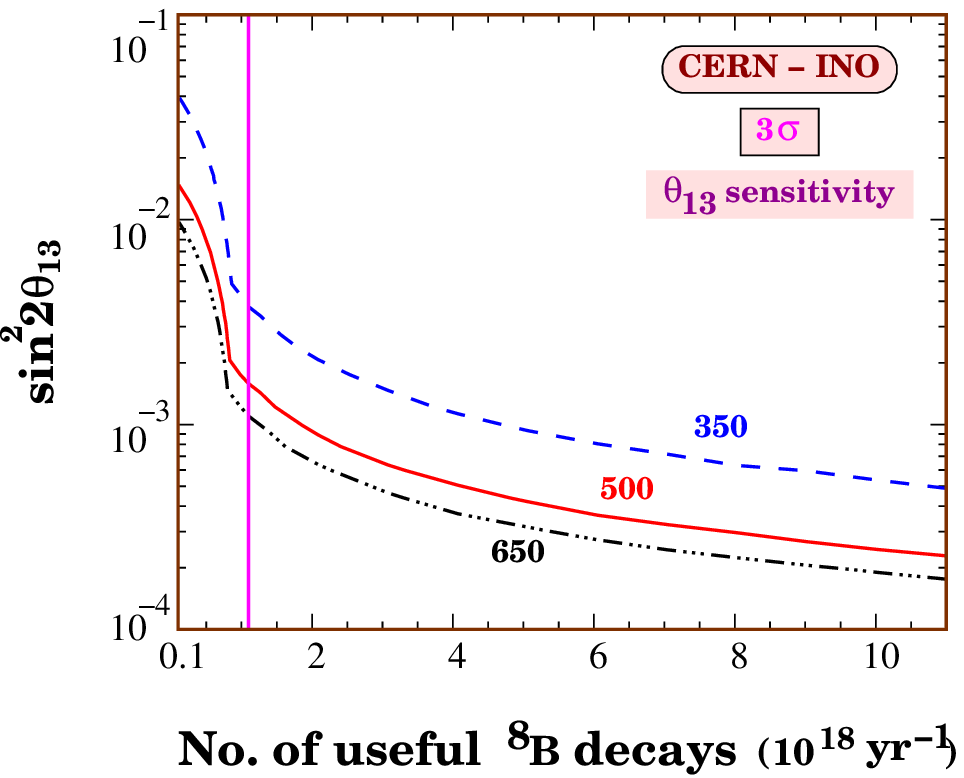}
\caption{\label{fig:senslum}
The variation of the experimental sensitivity on the 
number of useful ion decays in the straight sections of the 
storage ring. Left panel shows sensitivity to the 
mass hierarchy assuming NH to be true. Right panel shows the 
$\stch$ sensitivity reach. In both panels, the majenta solid
vertical line corresponds to the reference value used in the 
rest of the analysis.
}
\end{figure}
%

In Fig. \ref{fig:senshier} we show the minimum value of 
$\stch$ for which it would still be possible to 
reject the wrong hierarchy by this experimental set-up 
at the $3\sigma$ C.L.\footnote{We have given the 
details of our numerical analysis method in the Appendix.}. 
The left panel shows 
the hierarchy sensitivity 
when the normal hierarchy is true, while the right panel 
corresponds to the inverted hierarchy being true. 
We show this as a function of $\gamma$.  
In both panels we show by the red solid lines the sensitivity 
when we add neutrino and antineutrino data with the 
same value of the Lorentz boost shown in the $x$-axis. 
The blue dashed lines on the other hand correspond to the sensitivity 
expected when the neutrino beam runs with the $\gamma$ 
shown in the $x$-axis while the $\gamma$ for the 
antineutrino beam is scaled down by a factor of 
1.67.
We assume five years of running of the beta-beam in both polarities
and we have performed a full spectral analysis.
We note that using the combined neutrino and antineutrino beam 
running at the same value of $\gamma$ significantly enhances 
the hierarchy sensitivity of the experiment and the wrong 
hierarchy could be ruled out at $3\sigma$ for 
$\stch > 6.8 \times 10^{-4}$ ($\stch > 4.0 \times 10^{-4}$)
for $\gamma = 500$ if the normal 
(inverted)
hierarchy is true. 
This should be compared to the 
results from our earlier paper where 
we demonstrated that 
by using the 
beam with a single polarity, 
the wrong 
hierarchy can 
be rejected at the $3\sigma$ C.L. for 
$\stch > 9.8 \times 10^{-3}$ ($\stch > 8.2 \times 10^{-3}$) when 
the normal (inverted) 
hierarchy is true. Partial improvement in 
the hierarchy sensitivity comes 
from the fact that we have used spectral information here,
while in our earlier analysis we had considered just the total 
event rate measured in the detector. However, here the major improvement 
comes due to the addition of {\it both} the neutrino and antineutrino 
events, while in \cite{betaino} we had considered 
events due to {\it either} the neutrino (for true normal hierarchy)
{\it or} the antineutrino (for true inverted hierarchy) beam alone. 
Presence of both neutrino and antineutrino data simultaneously 
in the analysis restricts the fitted value of $\theta_{13}$ 
to be in a range very close 
to the assumed true value. 
For instance, for NH true, data corresponds to a large 
number of events for neutrinos and a small number of 
events for antineutrinos. When this is fitted with 
IH, we have a small number of events predicted for the 
neutrinos. In order to minimize the disparity between the 
data and prediction for neutrinos, 
the fit tends to drive $\theta_{13}$ 
to its largest allowed value. However, larger values of 
$\theta_{13}$ would give very large number of antineutrino 
events for IH and this would be in clear conflict with the data.
Therefore, the net advantage
of adding data from both neutrino and antineutrino 
channels is that 
one cannot artificially reduce the $\chi^2$ any longer 
by tinkering with $\theta_{13}$ in the fit. As a result,
the sensitivity of the experiment to 
mass hierarchy witnesses a substantial improvement.

We note from the plots that the hierarchy sensitivity falls when we 
use the scaled $\gamma$ option for the antineutrino beam. This is
particularly relevant 
when the true hierarchy is inverted and/or when 
$\gamma$ is low. 
Since scaling the $\gamma$ reduces it by a factor of 
1.67, the statistics for the antineutrinos fall by 
nearly a factor of 1.67 for this case and this reflects 
in the reduced hierarchy sensitivity of the experiment. 
Its impact when true hierarchy is inverted is  
more because in that case, the data 
corresponds to larger events for the antineutrinos and very 
small events for the neutrinos. The antineutrino events 
are therefore the driving force and 
an increase in their statistical uncertainty due to the scaled 
down $\gamma$ accentuates the adverse effect on the 
hierarchy sensitivity.
In the case of normal hierarchy, the events in the 
neutrino channel are the dominant factor and the 
role of the antineutrinos is only 
to prevent the $\theta_{13}$ 
values in the fit to run to very large values, as discussed 
before. 
As long as the antineutrino events have enough statistical power 
to restrict $\theta_{13}$ to values close to the 
true value at which the data was generated, the hierarchy sensitivity 
remains reasonably good. Therefore, for the 
normal hierarchy only for very low values of $\gamma$ 
the hierarchy sensitivity 
gets seriously affected by the Lorentz boost scaling. 

In Fig. \ref{fig:senshiertest} we show how the hierarchy 
sensitivity depends on diverse input factors. As in 
Fig. \ref{fig:senshier} we show the $3\sigma$
limit for $\stch$ as a function of $\gamma$ in all the 
four panels and we assume that normal hierarchy is true. 
The reference curve (red solid line) in all panels 
corresponds to the result obtained with a $\nue$ and 
$\anue$ beam with a spectral analysis.
The upper left 
hand panel shows the effect of changing the threshold energy  
of the detector. The sensitivity of the experiment 
is seen to remain  
almost stable against the variation of the threshold 
energy upto 4 GeV. 
Only for a threshold of 6 GeV and above the sensitivity falls, 
the lower $\gamma$ values getting more affected since 
they correspond to lower neutrino energies. 
In the lower left hand panel of the figure we 
show the effect of the chosen 
background fraction on the hierarchy 
sensitivity. The red solid line shows the sensitivity for our 
assumed background factor of $10^{-4}$ 
while the black dashed 
line shows the corresponding sensitivity when the
background rejection is poorer and we have 
a higher residual 
background fraction of $10^{-3}$. 
The upper right hand panel shows how our sensitivity 
increases by taking into account the spectral 
information of the events. It also shows 
how much improvement we get by combining the antineutrino 
data with the neutrino data. 
The black dashed line 
shows how the sensitivity falls when we use the total 
event rates instead of the events spectrum. The blue dashed-dotted
and green dashed-triple-dotted 
lines show the sensitivity expected from the neutrino 
data alone. The blue dashed-dotted 
line is for binned neutrino data while 
the green dashed-triple-dotted lines shows the sensitivity for the 
total event rate for neutrinos alone. We see that the effect of using the 
spectral information is only marginal when both neutrino 
and antineutrino are used, 
while the effect of combining 
the antineutrino data with the neutrino data 
on the sensitivity is huge. 
For the neutrino data alone, the sensitivity improves 
significantly when 
one uses the spectral information. 
In the lower right hand panel we show how the sensitivity of the 
experiment to hierarchy would get affected if 
we use a different profile for the 
Earth matter density instead of PREM. The red solid line is for 
earth density according to the PREM profile while  
the blue dotted and 
black dashed lines are when the matter density is 5\% lower 
and 5\% higher respectively than the density predicted 
by the PREM profile. When the density is higher (lower) 
the matter effects are higher (lower) and therefore the 
sensitivity improves (deteriorates).   

One crucial point that we have not stressed so far 
concerns the dependence of the detector performance on the 
{\it true value} of $\delta_{CP}$.  
All the earlier plots were presented assuming that 
$\delta_{CP}{\mbox {(true)}}=0$. 
At exactly the magic baseline, 
we expect the sensitivity of 
the experiment to be completely independent of $\delta_{CP}$. 
The CERN-INO distance of 7152 km is almost magical,
but it is not the exact magic baseline. 
Therefore, we do expect some remnant impact of $\delta_{CP}{\mbox {(true)}}$ 
on our results\footnote{Note that in all our results 
presented in this paper, we have 
fully marginalized over all the oscillation parameters in the fit, 
including $\delta_{CP}$.}. 
To show how our results get affected by 
$\dcpt$, we show in 
Fig. \ref{fig:senshiercploop} the hierarchy sensitivity just as 
in Fig. \ref{fig:senshier}, but here we show the full band corresponding 
to all values of $\dcpt$ from 0 to $2\pi$. As before, 
the left panel is for NH true while the the right panel is for IH true, 
and we have taken in the analysis the full spectral data 
for the neutrinos as well as the antineutrinos,  
with the same $\gamma$. The 
lower edge of this band shows the best possible scenario where
the experiment is most sensitive, while the 
upper edge shows the worst possible sensitivity.
The red solid lines in both panels show for comparison 
the hierarchy sensitivity 
corresponding to $\dcpt=0$, which we had presented in 
Fig. \ref{fig:senshier}. We note from the figure that 
the hierarchy sensitivity is nearly the 
best for $\dcpt=0$ when IH is true while if NH is 
true then it would give us almost the worst sensitivity. 
For NH (IH) as true the best possible sensitivity would be 
$\stch > 3.96 \times 10^{-4}$ ($\stch > 2.96 \times 10^{-4}$)
for $\gamma=650$ to be compared with $\stch > 5.51 \times 10^{-4}$
($\stch > 3.05 \times 10^{-4}$) when $\dcpt=0$.
Therefore, we conclude that if NH is true then 
it would not be unfair to
expect an 
even better hierarchy
sensitivity than what was reported in Fig.
\ref{fig:senshier}, while if IH is true then the best 
sensitivity will be returned for $\dcpt\simeq 0$. 

\begin{table}[t]
\begin{center}
\begin{tabular}{|c||c|c||c|c|} \hline
\multirow{2}{*}{$\gamma\backslash$N} 
& \multicolumn{2}{|c||}{{\rule[0mm]{0mm}{6mm}Mass Hierarchy (\sig)}} 
& \multicolumn{2}{|c|}{\rule[-3mm]{0mm}{6mm}{$\stch$ sensitivity (\sig)}}
\cr \cline{2-5} 
&  $1.1\times 10^{18}$ & $2.043\times 10^{18}$ & 
 $1.1\times 10^{18}$ & $2.043\times 10^{18}$ \cr
\hline\hline
350 & $1.3\times 10^{-3}$ & $9.3\times 10^{-4}$ &
 $3.8\times 10^{-3}$ & $2.3\times 10^{-3}$ \cr
650 & $5.6\times 10^{-4}$ & $4.1\times 10^{-4}$ &
 $1.1\times 10^{-3}$ & $7.3\times 10^{-4}$ \cr
\hline
\end{tabular}
\caption{\label{tab:compare}
Comparison of the variation of the detector sensitivity 
to mass hierarchy (columns 2 and 3) and $\stch$ sensitivity 
(columns 4 and 5) with $\gamma$ and N, the 
number of useful ion decays per 
year. 
}
\end{center}
\end{table}

We have noted from Figs. \ref{fig:eventshier} 
and \ref{fig:eventsth13} that 
the total number of events 
in the detector increases roughly linearly with $\gamma$, 
except for extremely long baselines. Increasing the 
number of ion decays per year will also bring about 
a simple linear increase in the statistics. 
It is therefore pertinent to make a 
fair comparison between the dependence of the 
mass hierarchy sensitivity to the Lorentz boost $\gamma$ and the 
number of useful ion decays in the ring\footnote{Note that 
this is also equivalent to increasing the total exposure time 
of the experiment. 
Both number of ion decays per year and exposure 
appear as a normalization factor for the event rate and hence 
increasing the number of ion decays by a factor $n$ keeping the
exposure same is equivalent to increasing the exposure by 
a factor $n$ keeping the number of ion decays per year fixed.
}.  
In the left panel of Fig. 
\ref{fig:senslum} we show the effect of increasing the 
number of ion decays on the hierarchy sensitivity\footnote{We assume
that the number of useful ion decays for both $^8B$ and $^8Li$ have
been scaled by the same factor. In the figure along the $x$-axis only
the $^8B$ numbers are shown.}. 
The plots exhibit the dependence of the sensitivity on
the number of useful ion decays per year for an exposure of five years, 
for three different values of $\gamma$.  
We have assumed the same Lorentz boost for the neutrino 
and antineutrino beams. We present in Table \ref{tab:compare}
the relative increase in the hierarchy sensitivity when 
we increase the $\gamma$ by a factor of 1.86 and compare it 
against the increase in the sensitivity when the number of 
ion decays are increased by the same factor. 
We note that while 
the hierarchy sensitivity improves by a factor of 
2.54 in going from $\gamma=350$ to 650 keeping the  
number of ion decays per year as $1.1\times 10^{18}$, it increases 
1.5-fold when we raise the number of ion decays per year from 
$1.1\times 10^{18}$ to $2.04\times 10^{18}$ keeping 
$\gamma=350$. However, we would like to stress 
that the improvement of the hierarchy sensitivity is not 
linear with either $\gamma$ or number of ion decays per year.
The crucial thing is that the behavior of the 
sensitivity dependence on both $\gamma$ and number of 
ion decays per year is very similar. 
It increases very fast initially and then 
comparatively flattens out.  

\section{Measurement of $\stch$}

\begin{figure}[t]
\includegraphics[width=8.0cm, height=7.0cm, angle=0]{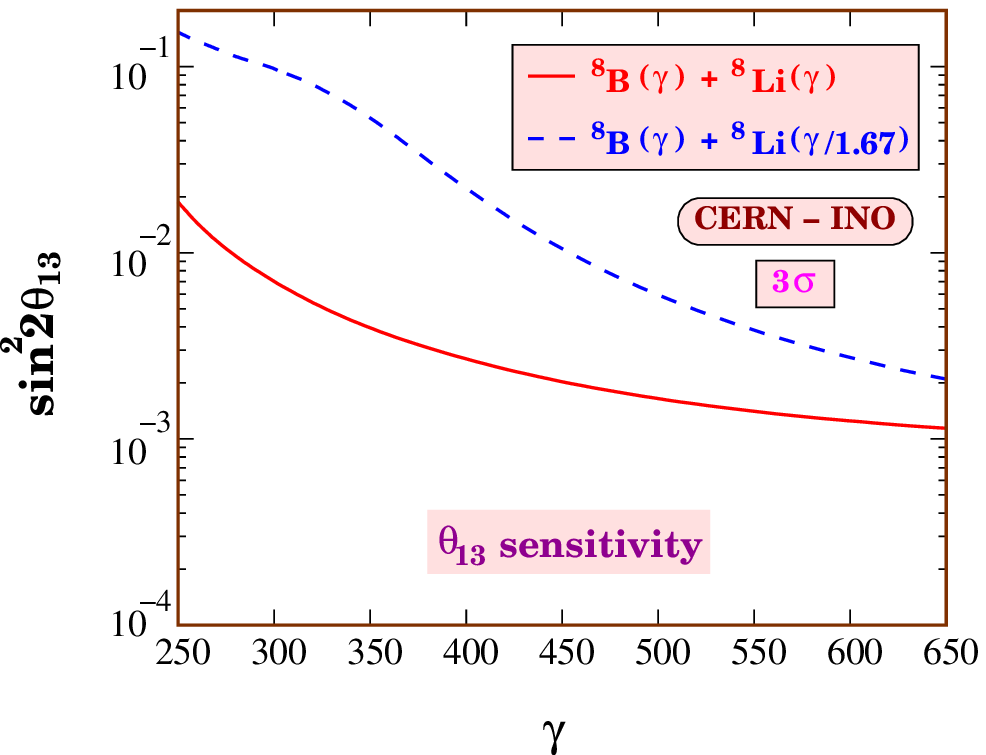}
\vglue -7.0cm \hglue 8.8cm
\includegraphics[width=8.0cm, height=7.0cm, angle=0]{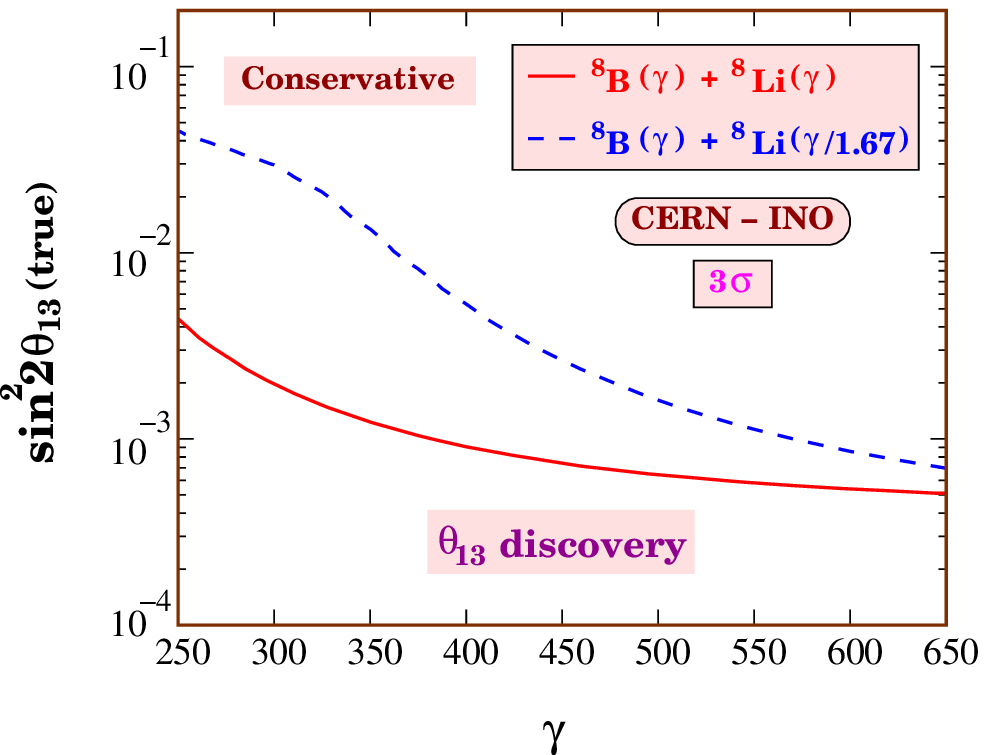}
\caption{\label{fig:sensth13}
Left panel shows 
the $3\sigma$ sensitivity limit for $\stch$. Right
panel shows 
the $3\sigma$ discovery reach for $\stcht$.
The red solid lines in the left and right panels 
show the sensitivity reach and discovery potential respectively,  
when the $\gamma$ is assumed to be the same 
for both the neutrino and the antineutrino beams. The blue dashed lines 
show the corresponding limits when the $\gamma$ for the 
$^8Li$ is scaled down by a factor of 1.67 with respect to the 
$\gamma$ of the neutrino beam, which is plotted in the $x$-axis.
}
\end{figure}
%

The CERN-INO beta-beam set-up is also expected to give 
very good sensitivity to the $\theta_{13}$ measurement.
In what follows, we will quantify our results in terms of 
three ``performance indicators'', 
\begin{enumerate}
\item  $\stch$ sensitivity reach,
\item $\stch$ discovery reach, 
\item $\stch$ precision.
\end{enumerate}
We give below a detailed description of our definition of these
performance indicators. For all results in this section we 
take into account the full event spectrum and combine
five yeats data 
from both the neutrino and antineutrino channels. 

\subsection{$\stch$ Sensitivity Reach}

We define the sensitivity reach of the CERN-INO 
beta-beam experiment 
as the upper limit on $\stch$ that can be put at the 
$3\sigma$ C.L.,   
in case no signal for $\theta_{13}$ 
driven oscillations is observed and the 
data is consistent with the null hypothesis. 
We simulate this situation in our analysis by 
generating the data at $\stcht=0$
and fitting it 
with some non-zero value of $\stch$ 
by means of the $\chi^2$ technique. 
In our fit we marginalize over
all the oscillation parameters including 
$\delta_{CP}$ and 
further marginalize over the 
mass hierarchy\footnote{Note that 
since $\stcht=0$, the data is independent of the mass hierarchy. 
However, since we allow for non-zero $\stch$ in the fit, the 
predicted event rates in our ``theory''
depend on the mass hierarchy.}  
and choose the value of 
$\stch$ for which the fit yields $\chi^2 = 9$.
The result is shown in the left
panel of Fig. \ref{fig:sensth13}, as a function of $\gamma$.
The red solid line shows the $\stch$ sensitivity when 
$\gamma$ is assumed to be the same 
for both the neutrino and the antineutrino beams.
The blue dashed lines 
shows the corresponding $3\sigma$ 
upper limit when $\gamma$ for the 
$^8Li$ is scaled down by a factor of 1.67 with respect to that
for the neutrino beam.
We can compare our results here with those obtained from 
our earlier analysis in \cite{betaino} where we had considered 
data from either the neutrino or the antineutrino channel 
alone when the hierarchy was normal and inverted, respectively.
We find that the $\stch$ we obtain from the combined neutrino 
and antineutrino data does not exhibit any marked improvement 
compared to that obtained in \cite{betaino}. 
This can be understood by the following reasoning.
In generating the data we have assumed that $\stcht=0$, 
which means that we have negligible events in both the 
neutrino as well as the antineutrino channels, irrespective 
of the mass hierarchy. When this data is fitted allowing for 
non-zero $\stch$, the neutrino (antineutrino) 
channel plays a dominating role
when NH (IH) is assumed in the fit. 
Therefore, 
the sensitivity we obtain assuming NH (IH) in our fit is 
similar to what we had in \cite{betaino} 
for the neutrino (antineutrino) channel alone. 
However, we reiterate that Fig. \ref{fig:sensth13} shows 
the $\stch$ sensitivity after marginalizing over hierarchy 
as well. In other words, the sensitivity shown in this figure
corresponds 
to the statistically weaker channel. 
For the case where we use same $\gamma$ for $^8B$ and 
$^8Li$, the neutrino channel is weaker  
since the event rate is about 1.5 times less 
than antineutrino events with the same $\gamma$.
On the other hand when we scale 
down the Lorentz boost for $^8Li$, the flux in the antineutrino 
channel goes down significantly and hence it becomes 
the statistically 
weaker channel as can be seen from Fig. \ref{fig:eventsbetaino}
and therefore 
the marginalized $\chi^2$ corresponds mainly to that from  
antineutrinos. Indeed one can check that the $\stch$ 
sensitivity that we exhibit by the blue dashed line for the scaled 
$\gamma$ case is comparable to what we had obtained for the 
antineutrino channel with IH and the corresponding lower 
$\gamma$. 

The dependence of the $\stch$ sensitivity on the number of 
useful radioactive ion decays per year 
in the straight section of the 
storage ring is shown in the right panel of Fig. \ref{fig:senslum}.
Here we have taken the same Lorentz boost for $^8B$ and $^8Li$
and we have shown the results for three fixed values of $\gamma$. 
The relative increase in the sensitivity by increasing 
$\gamma$ and/or the number of useful ion decays per year by the same 
factor is quantified in the last 
two columns of Table \ref{tab:compare}.

\subsection{$\stch$ Discovery Reach}

How good are our chances of observing a positive signal 
for oscillations and hence $\theta_{13}$ in the CERN-INO 
beta-beam set-up? We answer this question in 
terms of the parameter indicator which we call the ``discovery 
reach'' of the experiment for $\stch$. This is the minimum 
value of $\stcht$ that would give an unambiguous 
signal in the detector at $3\sigma$. 
To find this we simulate the 
data at some non-zero value of $\stcht$ and fit it by assuming that 
$\stch=0$, allowing all other oscillations parameters to take 
any possible value in order to return back the smallest value 
for the $\chi^2$. Note that since the fitted value of 
the mixing angle in this case 
always corresponds to 0, there is no 
need of any marginalizing over the hierarchy when fitting
the data. However, since the data here is generated at 
a non-zero value of $\stcht$, it depends on the 
true mass hierarchy. The discovery reach of the experiment 
is therefore expected to be dependent on the true mass hierarchy.
Likewise, while the value of $\delta_{CP}$ in the fit is 
inconsequential as $\stch=0$ in the fit, the data itself 
and hence the discovery reach, would depend on $\dcpt$. 
For each $\stcht$, 
we generate the data for all possible values of 
$\dcpt$ and for both the mass hierarchies. For each case, 
the data is then fitted assuming $\stch=0$
and marginalizing over the other oscillation parameters, 
returning a value of $\chi^2_{\rm min}$ for each data set. 
We choose the minimum amongst these $\chi^2_{\rm min}$ and 
find the value of $\stcht$ for which we could claim 
a signal in the detector at the $3\sigma$ C.L.  
In the right panel of 
Fig. \ref{fig:sensth13} we show this ``most conservative''\footnote{This 
is ``most conservative'' in the sense that no matter
what the choices of $\dcpt$ and the true neutrino mass ordering,
the $\theta_{13}$ discovery limit cannot be worse than the value presented.}
$\stch$ discovery reach 
of our experiment as a function of $\gamma$.
We assume 
equal $\gamma$ for both the ions for the red solid curve.
One can see that for $\gamma=650$, the most conservative
discovery reach is $\stcht=5.11 \times 10^{-4}$ while
if $\dcpt=0$ then this will be $\stcht=5.05 \times 10^{-4}$.
For the blue dashed line we assume that the $\gamma$ for 
$^8Li$ is scaled down by a factor of 1.67 compared to 
that for $^8B$, plotted on the $x$-axis. 
Since for 
same $\gamma$, neutrino is the statistically weaker channel, 
the red line mainly corresponds to what we expect for the 
true NH. For the scaled $\gamma$ case since the antineutrino 
channel becomes statistically weaker, the lower $\chi^2$ 
comes from this channel and the blue dashed line 
corresponds to what we expect for the 
true IH.

\subsection{$\stch$ Precision}
\begin{figure}[t]
\includegraphics[width=8.0cm, height=7.0cm]{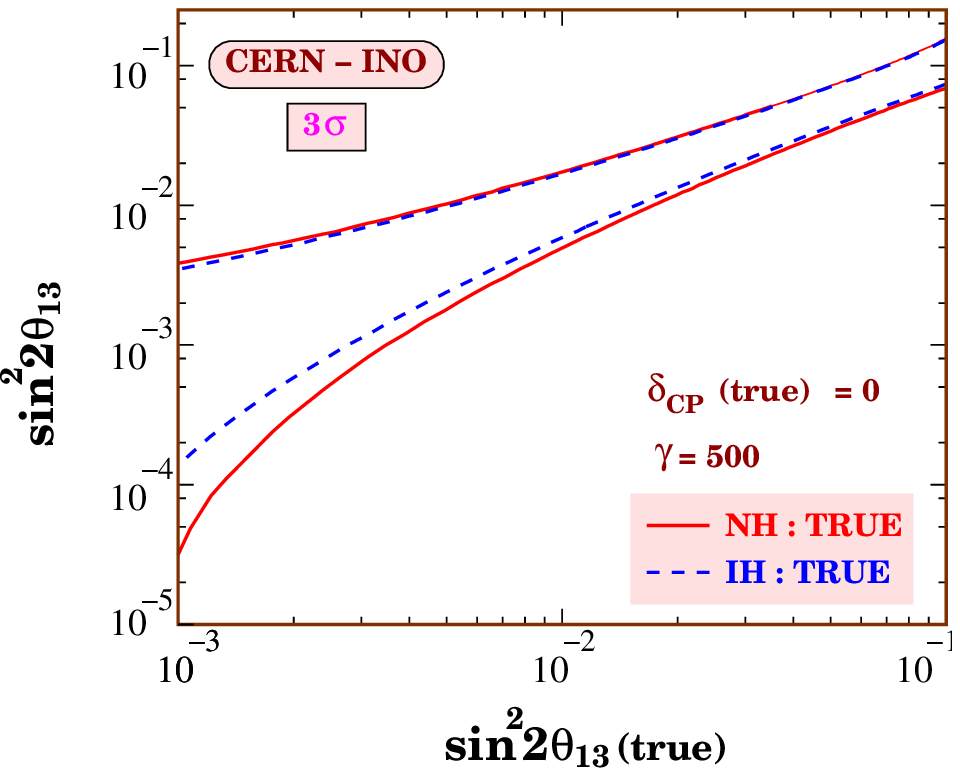}
\vglue -7.0cm \hglue 8.8cm
\includegraphics[width=8.0cm, height=7.0cm]{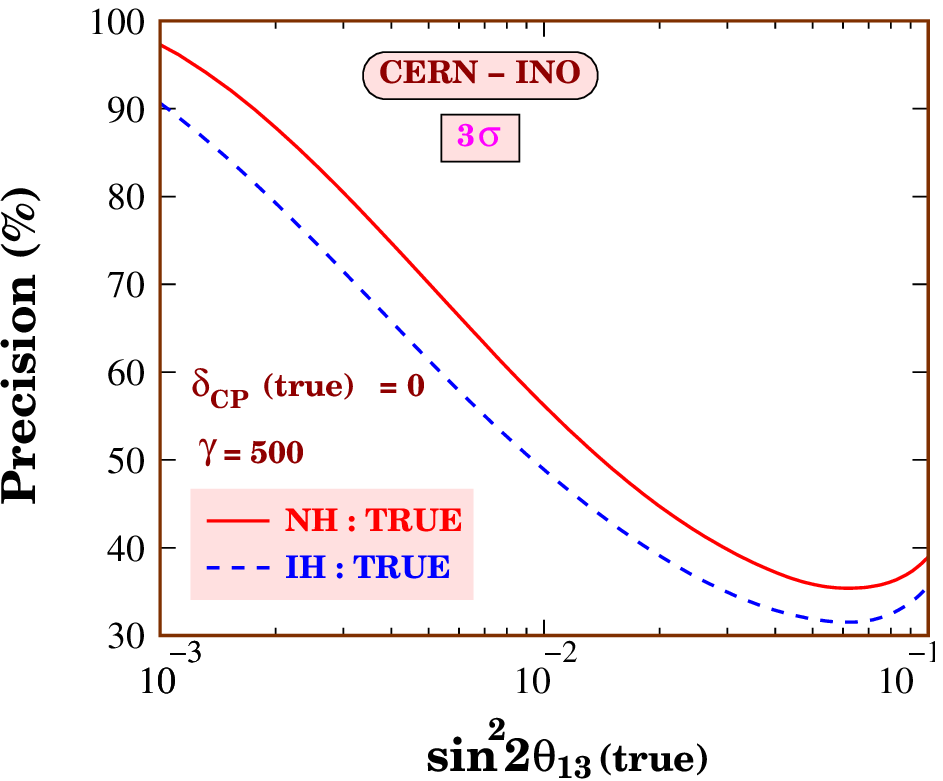}
\caption{\label{fig:precision}
The precision with which $\stch$ will be measured by 
the CERN-INO beta-beam experiment as a function of 
$\stcht$. Left panel shows the $3\sigma$ allowed range of $\stch$ 
while the right panel shows the precision defined in the text.
}
\end{figure}
%

In Fig. \ref{fig:precision} we show how {\it precisely} the mixing 
angle $\stch$ will be measured, if we observe a $\theta_{13}$
driven signal at the detector. 
The left panel depicts as a function 
of $\stcht$ the corresponding range of allowed values of 
$\stch$ at the $3\sigma$ C.L. 
We have assumed $\gamma=500$ and $\delta_{CP}=0$. 
The solid line is assuming NH to be true, while the dashed line 
is for IH true. Note that in the fit we always marginalize over the 
hierarchy and $\delta_{CP}$. The right panel shows the variable 
``precision'' which we define as 
\be
{\rm precision} = \frac{{(\stch)}_{max}-{(\stch)}_{min}}
{{(\stch)}_{max}+{(\stch)}_{min}} \times 100\,\%~,
\ee 
where ${(\stch)}_{max}$ and ${(\stch)}_{min}$ 
are the maximum and minimum allowed values of $\stch$ 
respectively 
at $3\sigma$.

\section{Conclusions}

Long baseline experiments which will use 
the golden $P_{e\mu}$ channel for 
determining the neutrino oscillation parameters face a serious 
threat from the menace of clone solutions due to the 
so-called parameter degeneracies. These degeneracies 
come in three forms: the $\delta_{CP}-\theta_{13}$ intrinsic 
degeneracy, the $ \delta_{CP}-sgn(\ma)$ degeneracy and 
the $\theta_{23}$ octant degeneracy, and necessarily result 
in degrading the sensitivity of the experiment. 
The CERN-INO near-magic distance of 7152 km offers the possibility of 
setting up an experiment at a
baseline where the $\delta_{CP}$ dependent terms almost drop 
out from the expression of the golden channel probability. 
Thus two out of the three degeneracies are evaded, providing  
a platform for clean measurement of $\theta_{13}$ and 
$sgn(\ma)$, two major players in our understanding of the 
origin of neutrino masses and mixing. 
A large magnetized 
iron calorimeter with a total mass of at least 50 kton is 
expected to be built soon at INO. It will be ideal for 
detecting multi-GeV $\numu$ and hence can be used as the far 
detector for a high energy beta-beam. 

In this paper we studied in detail the physics reach of the 
CERN-INO magical beta-beam set-up and extended our 
analysis presented in \cite{betaino}. 
Most importantly, we showed the impact of adding data from both the 
neutrino and antineutrino runs of the experiment.  
Combined data from both the neutrino and antineutrino polarities 
brings a major improvement in the hierarchy sensitivity 
of the experiment. 
We took into account 
the uncertainty due the solar parameters $\ms$ and $\sss$ 
and marginalized the $\chi^2$ over them. We also accounted 
for the uncertainty in the Earth matter density profile in 
our $\chi^2$ fit. We studied the impact of changing the 
energy threshold and the background rejection 
factor of the detector. We probed the importance of using the 
full spectral information on the final sensitivity of the experiment.
For $\gamma=650$, 
$\dcpt=0$ and true NH, 
the sensitivity to 
hierarchy determination at $3\sigma$ improves 
almost two orders of magnitude 
from 
$\stcht=1.15 \times 10^{-2}$ for the neutrino channel 
using the total rate only to
$\stcht=5.51 \times 10^{-4}$ 
when full spectral data from neutrino and antineutrino
channels are combined. Even though the effect of $\dcpt$ 
on the event rate of our experiment is expected to be small, 
there is some residual
dependence on it because the CERN-INO distance does not 
conform to the exact magic baseline. We studied the change in the 
hierarchy sensitivity due to the uncertainty in $\delta_{CP}$. 
It turns out that for $\gamma=650$ and with 
NH (IH) true, the best sensitivity to 
hierarchy determination corresponds to $\stcht =3.96 \times 10^{-4}$
($\stcht =2.96 \times 10^{-4}$), while the worst case is 
$\stcht =5.58 \times 10^{-4}$
($\stcht =3.59 \times 10^{-4}$).

We presented a detailed analysis of the potential of 
probing $\theta_{13}$ at this experiment. We defined and studied 
the $\theta_{13}$ reach in terms of three performance indicators:
the sensitivity reach, the discovery reach and the precision of 
$\stch$ measurement. The sensitivity reach is defined as the 
upper limit on $\stch$ that we would be able to impose in case the
data is consistent with null oscillations. At $3\sigma$ C.L. 
the sensitivity reach corresponds to 
$\stch=1.14 \times 10^{-3}$ 
for $\gamma=650$ and this is independent of the true hierarchy 
and $\dcpt$. 
The discovery reach is 
defined as the true value of the mixing angle for which we 
have an unambiguous oscillation signal in the detector. 
At $3\sigma$ C.L. 
the discovery reach corresponds to 
$\stcht=5.05 \times 10^{-4}$ ($\stcht=2.96 \times 10^{-4}$)
for $\gamma=650$, $\dcpt=0$ and NH (IH) true
while the most conservative limit irrespective of
$\dcpt$ and the true neutrino mass ordering is $\stcht=5.11 \times 10^{-4}$. 
We also presented the expected precision with which $\stch$ would be 
determined in this experiment for $\stcht > 10^{-3}$. 

Neutrino physics is in wait for the next great leap forward 
in the decade ahead. Beta-beams and an iron calorimeter 
detector at a very long baseline may well turn out,
as we have demonstrated, to be a key player in this 
endeavour.

\vglue 0.5cm
\noindent
{\Large{\bf Acknowledgments}}\vglue 0.3cm
\noindent
We are grateful to A. Samanta for providing us with the 
simulated number of atmospheric events expected in INO.
We thank W. Winter for helpful discussions, 
and acknowledge the HRI cluster facilities for computation. 
This work has been supported by the XIth Plan Neutrino Project
of the Harish-Chandra Research Institute.


\vglue 0.8cm
\noindent
{\Large{\bf Appendix: The Numerical Analysis}}
\vglue 0.3cm

In this appendix, we give more details of 
our numerical technique
through
which we statistically 
explore the physics potential of our set up.
The $\numu$ induced 
$\mu^-$ event spectrum at INO is estimated using
\be
N_{i} = T\, n_n\, f_{ID}\,\epsilon~  \int_0^{E_{\rm max}} dE
\int_{E_A^{\rm min}}^{E_A^{\rm max}}
dE_A \,\phi(E) \,\sigma_\numu(E) \,R(E,E_A)\, P_{e\mu}(E)
\label{eq:events}
\ee
where $N_i$ are the number of events in energy bin $i$ with 
lower energy limit $E_A^{\rm min}$ 
and upper  energy limit $E_A^{\rm max}$,
$T$ is the exposure time, 
$n_n$ are the number of target nucleons, $f_{ID}$ is the 
charge identification efficiency, $\epsilon$ is the detection 
efficiency, 
$\phi(E)$ is the  
beta-beam flux at INO, 
$\sigma_\numu$ is the detection cross section for $\numu$, 
$R(E,E_A)$ is the detector 
energy resolution function\footnote{
We assume a Gaussian resolution function with $\sigma=0.15E$.},
$E$ being the true energy of the incoming neutrino
and $E_A$ the measured energy of the muon. 
The corresponding 
$\mu^+$ spectrum due to $\anumu$ 
is given by replacing $\phi_\nue$ by $\phi_\anue$, $\sigma_\numu$ 
by $\sigma_\anumu$ and $P_{e\mu}$ by $P_{\bar e{\bar\mu}}$
in Eq. (\ref{eq:events}). 
We have used $f_{ID}=0.95$, $\epsilon=0.8$ and taken 
the neutrino-nucleon interaction cross sections from 
\cite{globes,Messier:1999kj,Paschos:2001np}.
In this paper we consider only the neutral current backgrounds
coming from the beam itself. We assume that 
the background can be rejected very efficiently by 
imposing suitable cuts such that only a fraction of $10^{-4}$
of these neutral current backgrounds survive.  
The background is assumed to have the same shape
as the signal. But one should keep in mind the fact that this shape
is not much of an issue since anyway the background is  
very small. 

For our statistical analysis we employ a $\chi^2$ 
function defined as
\be
\chi^2_{total} = \chi^2_{\nu_e \rightarrow \nu_{\mu}} 
               + \chi^2_{\bar\nu_e \rightarrow \bar\nu_{\mu}} 
               + \chi^2_{prior}~,
\label{eq:tot_chisq}
\ee
where the first term is the contribution from the 
neutrino channel, the second term comes from the antineutrino 
channel, while the last term comes from imposing priors 
on the oscillation parameters which we allow to vary freely 
in our fit and which we expect will be determined better from 
other experiments at the time when the data from the CERN-INO 
beta-beam set-up would be finally available. 
The $\chi^2$ for the neutrino channel is given by
\be
\chi^2_{\nu_e \rightarrow \nu_{\mu}} = min_{\xi_s, \xi_b}\left[2\sum^{n}_{i=1} 
(\tilde{y}_{i}-x_{i} - x_{i} \ln \frac{\tilde{y}_{i}}{x_{i}}) +
\xi_s^2 + \xi_b^2\right ]~.
\label{eq:chipull}
\ee
where $n$ is the total number of bins, 
\be
\tilde{y}_{i}(\{\omega\},\{\xi_s, \xi_b\}) = N^{th}_i(\{\omega\}) \left[
1+ \pi^s \xi_s \right] + 
N^{b}_i \left[1+ \pi^b \xi_b \right] ~,
\label{eq:rth}
\ee
$N^{th}_i(\{\omega\})$ given by Eq. (\ref{eq:events})
being the predicted number of 
events in the energy bin $i$ for a set of oscillation 
parameters $\omega$ and $N_i^b$ are the number of 
background events in bin $i$. The quantities 
$\pi^s$ and $\pi^b$ in Eq. (\ref{eq:rth}) are the systematical 
errors on signals and backgrounds respectively.
We have taken $\pi^s = 2.5\%$ and
$\pi^b = 5\%$. The quantities 
$\xi_s$ and $\xi_b$ are 
the ``pulls'' due to the systematical error
on signal and background respectively.
The data in Eq. (\ref{eq:chipull}) enters through the 
variables $x_i=N_i^{ex}+N_i^b$, where $ N_i^{ex}$ 
are the number of observed signal events in the detector and 
$N_i^b$ is the background, as mentioned earlier. We simulate  
the signal event spectrum using Eq. (\ref{eq:events})
for our assumed true values for the set of oscillation 
parameters. These assumed true values are given in 
the first column of Table \ref{tab:bench}. For 
$\stcht$, $\dcpt$ and the true hierarchy we use 
different options and mention them wherever applicable. 
In our $\chi^2$ fit we marginalize over {\it all} 
oscillation parameters, the Earth matter density, as well  
as the neutrino mass hierarchy, as applicable. We do this by allowing all 
of these to vary freely in the fit and picking the 
smallest value for the $\chi^2$ function. Of course we 
expect better determination of some of these parameters, 
which are poorly constrained by this experimental set-up. 
Therefore, we impose a ``prior'' on these parameters through the 
$\chi^2_{prior}$ given by
\be
\chi^2_{prior} &=&
\left (\frac{|\Delta m^2_{31}|-
|\mat|}{\sigma(|\Delta m^2_{31}|)} \right )^2 +
\left (\frac{\sta-\stat}{\sigma(\sta)} \right )^2\nonumber \\
&+&
\left (\frac{\Delta m^2_{21}-
\mst}{\sigma(\Delta m^2_{21})} \right )^2 +
\left (\frac{\sss-\ssst}{\sigma(\sss)} \right )^2\nonumber \\
&+&
\left (\frac{\rho-1}{\sigma(\rho)} \right )^2 ~.
\label{eq:prior}
\ee
where the $1\sigma$ error on these that we use are 
taken from \cite{huber10,solarprecision}
and are given in the left column of Table \ref{tab:bench}.  
In our computation, we have used a matter
profile inside the Earth 
with 24 layers. In Eq. (\ref{eq:prior}), $\rho$ is 
a constant number by
which the matter density of each layer has been scaled. 
The external information 
on $\rho$ is assumed to come from the study of the tomography of the earth
\cite{tomography}. In Eq. (\ref{eq:prior}), $\rho$ varies from
0.95 to 1.05 i.e., 5$\%$ fluctuation around 1. 

\begin{table}[t]
\begin{center}
\begin{tabular}{|l||l|}
\hline
& \\[-.5mm]
$|\Delta m^2_{31}{\rm (true)}| = 2.5 \times 10^{-3} \ {\rm eV}^2$ 
& $\sigma(\Delta m^2_{31})=1.5\%$ 
\\[2mm]
$\sin^2 2 \theta_{23}{\rm (true)} = 1.0$ 
& $\sigma(\sta)=1\%$ 
\\[2mm]
$\Delta m^2_{21}{\rm (true)} = 8.0 \times 10^{-5} \ {\rm eV}^2$ 
& $\sigma(\Delta m^2_{21})=2\%$ 
\\[2mm]
$\sin^2\theta_{12}{\rm (true)} = 0.31$ 
& $\sigma(\sss)=6\%$ 
\\[2mm]
$\rho{\rm (true)} = 1~{\rm (PREM)}$ & $\sigma(\rho)=5\%$\\[2mm]
\hline
\end{tabular}
\caption{\label{tab:bench}
Chosen benchmark values of oscillation parameters and their 
$1\sigma$ estimated error. The last row gives the corresponding 
values for the Earth matter density.
}
\end{center}
\end{table}

Note that in our definition of the $\chi^2$ function
given by Eq. (\ref{eq:tot_chisq}) and (\ref{eq:chipull}) 
we have assumed that the neutrino and antineutrino 
channel are completely uncorrelated, 
all the energy bins for a given channel 
are fully correlated, and 
$\xi_{s}$ and $\xi_{b}$ are fully uncorrelated.
We minimize the $\chi^2_{total}$ in two stages. First it 
is minimized with respect to $\xi_{s}$ and $\xi_{b}$ to 
get Eq. (\ref{eq:chipull}), and then with respect to the 
oscillation parameters ${\omega}$ to get the global best-fit.  
For minima with respect to $\xi_{s}$ and $\xi_{b}$, 
we require that
\be
\frac{\partial{\chi^2}}{\partial \xi_s} = 0 ~~{\rm and}~~ 
\frac{\partial{\chi^2}}{\partial \xi_b} = 0~.
\label{eq:minima}
\ee
From Eq. (\ref{eq:chipull}, \ref{eq:rth}, \ref{eq:minima})
we get,
\be
\left(
\begin{array}{cc}
a_{11} & a_{12} \\
a_{21} & a_{22} \end{array} \right)
\left(
\begin{array}{c}
\xi_{s} \\
\xi_{b} \end{array} \right)
=
\left(
\begin{array}{c}
c_{1} \\
c_{2} \end{array} \right)
\label{eq:matrix}
\ee
where,
\be
c_{1} = \sum^{n}_{i=1}
(\frac{x_{i} \pi^{s} N^{th}_{i}}{N^{th}_i + N^{b}_i} - \pi^{s} N^{th}_{i})~,
\nonumber \\
c_{2} = \sum^{n}_{i=1}
(\frac{x_{i} \pi^{b} N^{b}_{i}}{N^{th}_i + N^{b}_i} - \pi^{b} N^{b}_{i})~,
\nonumber \\
a_{11} = \sum^{n}_{i=1}
{\left[\frac{x_{i} (\pi^{s} N^{th}_{i})^{2}}{(N^{th}_i + N^{b}_i)^{2}} \right ]} + 1~,
\nonumber \\
a_{22} = \sum^{n}_{i=1}
{\left[\frac{x_{i} (\pi^{b} N^{b}_{i})^{2}}{(N^{th}_i + N^{b}_i)^{2}} \right ]} + 1~,
\nonumber \\
a_{12} = a_{21} = \sum^{n}_{i=1}
{\left[\frac{x_i N^{th}_{i} N^{b}_{i} \pi^{s} \pi^{b}}{(N^{th}_i + N^{b}_i)^{2}} \right ]}
\label{eq:coeffi}
\ee 
Using Eq. (\ref{eq:matrix} and \ref{eq:coeffi}),
we calculate the values of $\xi_{s}$ and $\xi_{b}$ and then we use
these values to calculate $\chi^2_{\nu_e \rightarrow \nu_{\mu}}$.
In a similar fashion, we estimate 
$\chi^2_{\bar\nu_e \rightarrow \bar\nu_{\mu}}$
to obtain the $\chi^2_{total}$.



\begin{thebibliography}{99}

\bibitem{solar}
B.~T.~Cleveland {\it et al.},
Astrophys.\ J.\  {\bf 496}, 505 (1998);
%
J.~N.~Abdurashitov {\it et al.}  [SAGE Collaboration],
J.\ Exp.\ Theor.\ Phys.\  {\bf 95}, 181 (2002)
[Zh.\ Eksp.\ Teor.\ Fiz.\  {\bf 122}, 211 (2002)];
%
W.~Hampel {\it et al.}  [GALLEX Collaboration],
Phys.\ Lett.\ B {\bf 447}, 127 (1999); 
S.~Fukuda {\it et al.}  [Super-Kamiokande Collaboration],
Phys.\ Lett.\ B {\bf 539}, 179 (2002);
%
B.~Aharmim {\it et al.}  [SNO Collaboration],
Phys.\ Rev.\ C {\bf 72}, 055502 (2005).

\bibitem{borex}
C.~Arpesella {\it et al.},
  [Borexino Collaboration],
  arXiv:0708.2251 [astro-ph].

\bibitem{kl}
K.~Eguchi {\it et al.}, 
  [KamLAND Collaboration],
Phys.\ Rev.\ Lett.\  {\bf 90}, 021802 (2003);
%
T.~Araki {\it et al.}  [KamLAND Collaboration],
Phys.\ Rev.\ Lett.\  {\bf 94}, 081801 (2005).

\bibitem{kltalk}
I. Shimizu, talk at 10th International Conference on 
Topics in Astroparticle and Underground Physics, TAUP 2007.

\bibitem{limits}
M.~Maltoni, 
T.~Schwetz, M.~A.~Tortola and J.~W.~F.~Valle,
New J.\ Phys.\  {\bf 6}, 122 (2004), hep-ph/0405172 v5;
%
  S.~Choubey,
  arXiv:hep-ph/0509217;
%
  S.~Goswami,
  Int.\ J.\ Mod.\ Phys.\ A {\bf 21}, 1901 (2006);
%
  A.~Bandyopadhyay, 
S.~Choubey, S.~Goswami, S.~T.~Petcov and D.~P.~Roy,
  Phys.\ Lett.\ B {\bf 608}, 115 (2005);
%
  G.~L.~Fogli {\it et al.}, 
  Prog.\ Part.\ Nucl.\ Phys.\  {\bf 57}, 742 (2006).

\bibitem{atm}
  Y.~Ashie {\it et al.}  [Super-Kamiokande Collaboration],
  Phys.\ Rev.\ D {\bf 71}, 112005 (2005).

\bibitem{k2k}
E.~Aliu {\it et al.}  [K2K Collaboration],
  Phys.\ Rev.\ Lett.\  {\bf 94}, 081802 (2005). 

\bibitem{minos}
D. G. Michael {\it et al.}, [MINOS Collaboration],
  Phys.\ Rev.\ Lett.\  {\bf 97}, 191801 (2006).

\bibitem{0vbbus}
  S.~Pascoli, S.~T.~Petcov and T.~Schwetz,
  Nucl.\ Phys.\  B {\bf 734}, 24 (2006);
%
  S.~Choubey and W.~Rodejohann,
  Phys.\ Rev.\  D {\bf 72}, 033016 (2005).

\bibitem{chooz}
M.~Apollonio {\it et al.},
Eur.\ Phys.\ J.\ C {\bf 27}, 331 (2003).

\bibitem{msw1}
  L.~Wolfenstein,
  Phys.\ Rev.\ D {\bf 17}, 2369 (1978).

\bibitem{msw2}
  S.~P.~Mikheev and A.~Y.~Smirnov,
  Sov.\ J.\ Nucl.\ Phys.\  {\bf 42}, 913 (1985)
  [Yad.\ Fiz.\  {\bf 42}, 1441 (1985)];
%
  S.~P.~Mikheev and A.~Y.~Smirnov,
  Nuovo Cim.\ C {\bf 9}, 17 (1986).

\bibitem{msw3}
  V.~D.~Barger, K.~Whisnant, S.~Pakvasa and R.~J.~N.~Phillips,
  Phys.\ Rev.\ D {\bf 22}, 2718 (1980).

\bibitem{t2k}
  Y.~Itow {\it et al.},
  arXiv:hep-ex/0106019.


\bibitem{nova}
  D.~S.~Ayres {\it et al.}  [NOvA Collaboration],
  arXiv:hep-ex/0503053.

\bibitem{huber10}
  P.~Huber,
M.~Lindner, M.~Rolinec, T.~Schwetz and W.~Winter,
  Phys.\ Rev.\ D {\bf 70}, 073014 (2004)
and referecences therein.

\bibitem{iss}
http://www.hep.ph.ic.ac.uk/iss/

\bibitem{paper1}
  S.~K.~Agarwalla, A.~Raychaudhuri and A.~Samanta,
  Phys.\ Lett.\ B {\bf 629}, 33 (2005).

\bibitem{betaino}
  S.~K.~Agarwalla, S.~Choubey and A.~Raychaudhuri,
  Nucl.\ Phys.\  B {\bf 771}, 1 (2007).

\bibitem{ino}
  M.~S.~Athar {\it et al.}  [INO Collaboration],
 A Report of the INO Feasibility Study,\\
{http://www.imsc.res.in/~ino/OpenReports/INOReport.pdf}
%

\bibitem{zucc}
P.~Zucchelli,
Phys.\ Lett.\ B {\bf 532}, 166 (2002).

\bibitem{cernmemphys}
  J.~E.~Campagne, M.~Maltoni, M.~Mezzetto and T.~Schwetz,
  JHEP {\bf 0704}, 003 (2007).

\bibitem{intrinsic}
  J.~Burguet-Castell, 
M.~B.~Gavela, J.~J.~G\'{o}mez-Cadenas, P.~Hernandez and O.~Mena,
  Nucl.\ Phys.\ B {\bf 608}, 301 (2001).

\bibitem{minadeg}
  H.~Minakata and H.~Nunokawa,
  JHEP {\bf 0110}, 001 (2001).

\bibitem{th23octant}
  G.~L.~Fogli and E.~Lisi,
  Phys.\ Rev.\ D {\bf 54}, 3667 (1996).

\bibitem{eight}
  V.~Barger, D.~Marfatia and K.~Whisnant,
  Phys.\ Rev.\ D {\bf 65}, 073023 (2002).

\bibitem{magic}
  P.~Huber and W.~Winter,
  Phys.\ Rev.\ D {\bf 68}, 037301 (2003).

\bibitem{magic2}
  A.~Y.~Smirnov,
  arXiv:hep-ph/0610198.

\bibitem{petcov}
  M.~Freund, M.~Lindner, S.~T.~Petcov and A.~Romanino,
  Nucl.\ Phys.\  B {\bf 578}, 27 (2000).


\bibitem{prem}
  A.~M.~Dziewonski and D.~L.~Anderson,
  Phys.\ Earth Planet.\ Interiors {\bf 25}, 297 (1981);\\
S.~V.~Panasyuk, Reference Earth Model (REM) webpage,\\
 http://cfauves5.harvrd.edu/lana/rem/index.html.

\bibitem{betaoptim}
  P.~Huber, M.~Lindner, M.~Rolinec and W.~Winter,
  Phys.\ Rev.\  D {\bf 73}, 053002 (2006).

\bibitem{bc}
  J.~Burguet-Castell, D.~Casper, E.~Couce, J.~J.~G\'{o}mez-Cadenas and P.~Hernandez,
  Nucl.\ Phys.\  B {\bf 725}, 306 (2005);
%
  J.~Burguet-Castell, D.~Casper, J.~J.~G\'{o}mez-Cadenas, P.~Hernandez and F.~Sanchez,
  Nucl.\ Phys.\  B {\bf 695}, 217 (2004).

\bibitem{volpe}
C.~Volpe,
  J.\ Phys.\ G {\bf 34}, R1 (2007).

\bibitem{oldpapers}
  M.~Mezzetto,
  J.\ Phys.\ G {\bf 29}, 1771 (2003);
%
  M.~Mezzetto,
  Nucl.\ Phys.\ Proc.\ Suppl.\  {\bf 143}, 309 (2005);
%
  M.~Mezzetto,
  Nucl.\ Phys.\ Proc.\ Suppl.\  {\bf 155}, 214 (2006).
%

\bibitem{doninibeta}
  A.~Donini, E.~Fernandez-Martinez, P.~Migliozzi, S.~Rigolin and L.~Scotto Lavina,
  Nucl.\ Phys.\  B {\bf 710}, 402 (2005);
%
  A.~Donini, 
E.~Fernandez, P.~Migliozzi, S.~Rigolin, L.~Scotto Lavina, T.~Tabarelli de Fatis and F.~Terranova,
  arXiv:hep-ph/0511134;
%
  A.~Donini, E.~Fernandez-Martinez, P.~Migliozzi, S.~Rigolin, L.~Scotto Lavina, T.~Tabarelli de Fatis and F.~Terranova,
  Eur.\ Phys.\ J.\  C {\bf 48}, 787 (2006).

\bibitem{rubbia}
  C.~Rubbia, A.~Ferrari, Y.~Kadi and V.~Vlachoudis,
  Nucl.\ Instrum.\ Meth.\ A {\bf 568}, 475 (2006);
  C.~Rubbia,
  arXiv:hep-ph/0609235.

\bibitem{mori}
 Y.~Mori, Nucl.\ Instrum.\ Meth.\ A {\bf 562}, 591 (2006).

\bibitem{doninialter}
  A.~Donini and E.~Fernandez-Martinez,
  Phys.\ Lett.\ B {\bf 641}, 432 (2006).

\bibitem{rparity}
  R.~Adhikari, S.~K.~Agarwalla and A.~Raychaudhuri,
  Phys.\ Lett.\ B {\bf 642}, 111 (2006);
%
  S.~K.~Agarwalla, S.~Rakshit and A.~Raychaudhuri,
  Phys.\ Lett.\  B {\bf 647}, 380 (2007).

\bibitem{olga}
  A.~Jansson, O.~Mena, S.~Parke and N.~Saoulidou,
  arXiv:0711.1075 [hep-ph].

\bibitem{golden}
  A.~Cervera, A.~Donini, M.~B.~Gavela, J.~J.~G\'{o}mez-Cadenas, P.~Hernandez, O.~Mena and S.~Rigolin,
  Nucl.\ Phys.\ B {\bf 579}, 17 (2000)
  [Erratum-ibid.\ B {\bf 593}, 731 (2001)].

\bibitem{freund}
  M.~Freund, P.~Huber and M.~Lindner,
  Nucl.\ Phys.\  B {\bf 615}, 331 (2001).

\bibitem{gandhi}
  R.~Gandhi, P.~Ghoshal, S.~Goswami, P.~Mehta and S.~Uma Sankar,
  Phys.\ Rev.\ Lett.\  {\bf 94}, 051801 (2005);
%
  R.~Gandhi, P.~Ghoshal, S.~Goswami, P.~Mehta and S.~Uma Sankar,
  Phys.\ Rev.\ D {\bf 73}, 053001 (2006).

\bibitem{pee}
  S.~K.~Agarwalla, S.~Choubey, S.~Goswami and A.~Raychaudhuri,
  Phys.\ Rev.\  D {\bf 75}, 097302 (2007).


\bibitem{mind}
A. Cervera, talk at NuFact07.

\bibitem{globes}
  P.~Huber, M.~Lindner and W.~Winter,
  Comput.\ Phys.\ Commun.\  {\bf 167}, 195 (2005).

\bibitem{Messier:1999kj}
  M.~D.~Messier,
PhD thesis, UMI-99-23965.

\bibitem{Paschos:2001np}
  E.~A.~Paschos and J.~Y.~Yu,
  Phys.\ Rev.\ D {\bf 65}, 033002 (2002).


\bibitem{solarprecision}
See for example,  
A.~Bandyopadhyay {\it et al.}, 
  Phys.\ Rev.\ D {\bf 72}, 033013 (2005);
  J.~N.~Bahcall and C.~Pena-Garay,
  JHEP {\bf 0311}, 004 (2003).

\bibitem{tomography}
R.~J.~Geller and T.~Hara,
Nucl.\ Instrum.\ Meth.\  A {\bf 503}, 187 (2001).

\end{thebibliography}
\end{document}